\documentclass[aps,prx,twocolumn,superscriptaddress,showpacs]{revtex4-1}

\usepackage{epsfig,amsmath}
\usepackage{subfigure}
\usepackage{graphicx}% Include figure files
\usepackage{dcolumn}% Align table columns on decimal point
\usepackage{stmaryrd}
\usepackage{mathrsfs}
\usepackage{pifont}
\usepackage{amsthm}
\usepackage{amssymb}
\usepackage{bm}
\usepackage{latexsym}
\usepackage[colorlinks=true,linkcolor=blue,citecolor=blue]{hyperref}
\usepackage{color}
\usepackage{bbold}
\usepackage{subfigure}

\newcommand{\la}{\langle}
\newcommand{\ra}{\rangle}

\newcommand{\ti}{\tilde}

\newcommand{\ga}{\gamma}
\newcommand{\Ga}{\Gamma}
\newcommand{\ka}{\kappa}
\newcommand{\da}{\dagger}
\newcommand{\De}{\Delta}
\newcommand{\al}{\alpha}
\newcommand{\si}{\sigma}

\newcommand{\om}{\omega}
\newcommand{\Om}{\Omega}
\newcommand{\de}{\delta}

\newcommand{\pa}{\partial}
\newcommand{\map}{\mathcal{P}}
\newcommand{\maq}{\mathcal{Q}}
\newcommand{\mai}{\mathcal{I}}

\newcommand{\bes} {\begin{subequations}}
\newcommand{\ees} {\end{subequations}}
\newcommand{\bea} {\begin{eqnarray}}
\newcommand{\eea} {\end{eqnarray}}
\newcommand{\beq} {\begin{equation}}
\newcommand{\eeq} {\end{equation}}

\newcommand{\ket}[1]{ | #1\rangle}

\newcommand{\ketbra}[2]{|#1\rangle\langle #2|}
\newcommand{\kett}[1]{ | #1\rangle\rangle}
\newcommand{\braa}[1]{\langle\langle #1 | }
\newcommand{\mc}[1]{\mathcal #1}
\newcommand{\ignore}[1]{}

\def\jpb#1{{ J.\ Phys.\ B} {\bf#1}}
\def\jpa#1{{ J.\ Phys.\ A} {\bf#1}}
\def\pra#1{{ Phys.\ Rev. A\/} {\bf#1}}
\def\prb#1{{ Phys.\ Rev. B\/} {\bf#1}}
\def\pre#1{{ Phys.\ Rev. E\/} {\bf#1}}
\def\prx#1{{ Phys.\ Rev. X\/} {\bf#1}}
\def\prl#1{{ Phys.\ Rev.\ Lett.} {\bf#1}}

\def\sci#1{{ Science} {\bf#1}}

\def\pla#1{{ Phys.\ Lett. A\/} {\bf#1}}
\def\rmp#1{{ Rev. \ Mod. \ Phys.} {\bf#1}}

\def\njp#1{{ New \ J. \ Phys.} {\bf#1}}

\begin{document}

\title{Eigenstate Tracking in Open Quantum Systems}

\author{Jun Jing}
\affiliation{Institute of Atomic and Molecular Physics and \\ Jilin Provincial Key Laboratory of Applied Atomic and Molecular Spectroscopy, Jilin University, Changchun 130012, Jilin, China.}
\affiliation{Department of Theoretical Physics and History of Science, The Basque Country University (EHU/UPV), PO Box 644, 48080 Bilbao, and Ikerbasque, Basque Foundation for Science, 48011 Bilbao, Spain.}
\author{Marcelo S. Sarandy}
\affiliation{Instituto de F\'{\i}sica, Universidade Federal Fluminense,  Campus da Praia Vermelha, 24210-346, Niter\'oi, RJ, Brazil.}
\affiliation{Departments of Electrical Engineering, Chemistry and Physics, and Center for Quantum Information Science \& Technology, University of Southern California, Los Angeles, California 90089, USA.}
\author{Daniel A. Lidar}
\affiliation{Departments of Electrical Engineering, Chemistry and Physics, and Center for Quantum Information Science \& Technology, University of Southern California, Los Angeles, California 90089, USA.}
\author{Da-Wei Luo}
\affiliation{Department of Theoretical Physics and History of Science, The Basque Country University (EHU/UPV), PO Box 644, 48080 Bilbao, and Ikerbasque, Basque Foundation for Science, 48011 Bilbao, Spain.}
%\affiliation{Beijing Computational Science Research Center, Beijing 100084, China.}
\author{Lian-Ao Wu}
\email{Corresponding author: lianaowu@gmail.com}
\affiliation{Department of Theoretical Physics and History of Science, The Basque Country University (EHU/UPV), PO Box 644, 48080 Bilbao, and Ikerbasque, Basque Foundation for Science, 48011 Bilbao, Spain.}

\date{\today}

\begin{abstract}
Keeping a quantum system in a given instantaneous eigenstate is a control problem with numerous applications, e.g., in quantum information processing. The problem is even more challenging in the setting of open quantum systems, where environment-mediated transitions introduce additional decoherence channels. Adiabatic passage is a well established solution, but requires a sufficiently slow evolution time that is dictated by the adiabatic theorem.
%An alternative is the use of assistant (counter-diabatic) Hamiltonians to achieve shortcuts to adiabaticity, but this typically results in highly non-local interactions and may be vulnerable to fluctuations in both system and environment.
Here we develop a systematic projection theory formulation for the transitionless evolution of general open quantum systems described by time-local master equations. We derive a time-convolutionless dynamical equation for the target instantaneous eigenstate of a given time-dependent Hamiltonian. A transitionless dynamics then arises in terms of a competition between the average Hamiltonian gap and the decoherence rate, which implies optimal adiabaticity timescales. We show how eigenstate tracking can be accomplished via control pulses, without explicitly incorporating counter-diabatic driving, thus offering an alternative route to shortcuts to adiabaticity. We examine rectangular pulses, chaotic signals, and white noise, and find that, remarkably, the effectiveness of eigenstate tracking hardly depends on the details of the control functions. In all cases the control protocol keeps the system in the desired instantaneous eigenstate throughout the entire evolution, along an accelerated adiabatic path.
\end{abstract}


\maketitle

\section{Introduction}

Tracking of an eigenstate of a Hamiltonian, in particular the ground state, is a protocol of great interest in quantum control~\cite{Brif:2010fu}, with numerous applications, e.g., in quantum information processing. The best known such tracking protocol is the adiabatic theorem of quantum mechanics~\cite{Born:28,Messiah:book}, which states that a system that is initially prepared in an eigenstate of a time-dependent Hamiltonian $H(t)$ will evolve to the corresponding instantaneous eigenstate at a later time $T$ provided that $H(t)$ varies smoothly and that $T$ is much larger than (some power of) the relevant minimal inverse eigenenergy gap of $H(t)$~\cite{Kato:50,Avron:87,Jansen:07}. Applications of eigenstate tracking have proliferated, covering research fields such as adiabatic quantum computation and quantum annealing~\cite{kadowaki_quantum_1998,farhi_quantum_2001,childs_robustness_2001,
aharonov_adiabatic_2007,Matsuda:2009uq,speedup}, holonomic quantum computation~\cite{HQC,Duan:2001ff,Oreshkov:2009bl}, adiabatic passage~\cite{Oreg,Bergmann,Kral,PhysRevA.87.022332}, adiabatic gates~\cite{Bacon:2013qy,Hen:15}, many-body state preparation~\cite{Hamma:2008jk,PhysRevB.91.134303}, and quantum phase transitions~\cite{dziarmaga_dynamics_2010,PhysRevA.82.012321}, to name a few. The adiabatic theorem can be viewed as providing a passive protocol for eigenstate tracking, where the main control knob is the total evolution time $T$. More accurate tracking can be achieved within the adiabatic framework by changing $H(t)$ more slowly near avoided crossings~\cite{Roland:2002ul,PhysRevLett.103.080502} or by imposing smooth boundary conditions on $H(t)$ and its derivatives~\cite{Garrido:62,Nenciu:81,lidar:102106,Wiebe:12,Ge:2015wo}, but the fact remains that the smaller the gap the longer is the time $T$ required for the system to track an eigenstate. This has motivated the investigation of methods to accelerate adiabaticity, such as the transitionless-tracking algorithm~\cite{Rice1,Rice2,Berry:2009kx}. In this reverse engineering method, an ``assistant Hamiltonian", or ``counter-diabatic driving" term, built from the instantaneous eigenstates of the original system Hamiltonian $H(t)$ is introduced in order to completely cancel the off-diagonal terms of $H(t)$ written in the adiabatic frame~\cite{Avron-assisted-comment}.
%%%
%\footnote{This assistant Hamiltonian is in fact identical to the adiabatic generator $\frac{i}{T}[P'(s),P(s)]$, where $s=t/T$ and $P(s)$ is the spectral projection on the appropriate energy band of $H(s)$, introduced first by Avron \textit{et al.} \cite{Avron:87}. This formulation is basis-independent.}.
%%%
When the counter-diabatic term is included, the full Hamiltonian ``superadiabatically" drives the system along the instantaneous eigenstate of the original $H(t)$ towards the target state, providing a shortcut to adiabaticity ~\cite{Torrontegui:13,Deffner:2014qv,Papoular,Santos:2015fk,Santos:2016uq}, albeit at the price of highly non-local interactions when applied to quantum many-body systems~\cite{Campo:2012il,Campo:2013ix}.

These theoretical results were developed in the context of closed quantum systems, evolving unitarily in the absence of an environment, or bath. Despite the fact that experiments implementing the superadiabatic protocol have already been reported~\cite{Bason:2012vh,Zhang:2013ak}, Hamiltonian eigenstate tracking is much less developed in the context of realistic, open quantum systems. Pertinent studies include an analysis of the effect of control noise using the Lewis-Riesenfeld invariant formalism~\cite{Ruschhaupt:2012hi,Jing13}, and superadiabatic protocols for improving the efficiency of heat engines or accelerating cooling~\cite{Hoffmann:2011ts,Deng:2013xu,Campo:2014rf,Stefanatos:2014bl}. Most relevant to our setting is a formal treatment of transitionless dynamics in open systems reported by Vacanti \textit{et al.}~\cite{Vacanti:2014qy}. This work builds on the Jordan block approach to adiabaticity in open system, wherein decoupling of Jordan blocks of the Lindbladian superoperator is identified with adiabatic evolution~\cite{PhysRevA.71.012331}. Transitions between Jordan blocks are suppressed in Ref.~\cite{Vacanti:2014qy} by adding an appropriate counter-diabatic driving term. This term, in general, requires quantum channel engineering, a highly non-trivial task.

Here, we propose a new approach to transitionless evolutions in open quantum systems, as well as a systematic method to accelerate adiabatic paths without explicitly incorporating any counter-diabatic driving, thus circumventing the problem of highly non-local interactions associated with the latter. Our results are applicable to general open quantum systems described by time-local master equations. We do not invoke the Jordan blocks approach; rather, we build on the standard notion of adiabatic evolution as being represented by decoupled Hamiltonian eigensubspaces or eigenstates. More specifically, using the Feshbach P-Q partitioning procedure~\cite{Wu09} and the Nakajima-Zwanzig projection technique~\cite{Breuer:book}, we derive a time-convolutionless (TCL) equation governing the population dynamics of an arbitrary Hamiltonian eigenstate subject to open system evolution [Eq.~\eqref{eq:TCL-result}]. This equation provides a general condition for eigenstate tracking in open quantum systems. Adiabatic perturbation theory (an expansion in powers of $1/T$) and a weak coupling expansion allow us to simplify the result into a form that lends itself to an interpretation in terms of diabatic or bath-induced transitions [Eq.~\eqref{de-open}], which can be suppressed using a control protocol. We emphasize that the approach adopted here constitutes a novel strategy towards the study of transitionless evolutions in open systems. Previous approaches to open system adiabaticity focused on Jordan block decoupling~\cite{PhysRevA.71.012331,PhysRevLett.95.250503}, the weak coupling limit~\cite{PhysRevA.72.022328}, coupling to an ancilla~\cite{0953-4075-40-2-004}, zero temperature~\cite{Pekola:2010oj}, Markovian evolution~\cite{ABLZ:12-SI}, or convergence to the instantaneous steady state of the Lindbladian~\cite{joye_general_2007,oreshkov_adiabatic_2010,
Avron:2012tv,Venuti:2015kq,salem_quasi-static_2007}. In contrast, our approach directly establishes conditions to keep the system in an eigenstate of the original Hamiltonian, the only assumption being that the system's evolution is described by a time-local master equation. It recovers the standard closed-system adiabatic theorem as a special case.

We illustrate our framework using examples involving the open system dynamics of a qubit coupled to various environments and subject to various control protocols. An important conclusion that emerges from these examples is that the condition for transitionless open system dynamics involves a competition between the average gap of the Hamiltonian superoperator and the decoherence rate, with the former favoring a long evolution time and the latter favoring a short evolution time. This interplay is reflected in a damped oscillatory behavior of the eigenstate fidelity as a function of the evolution time, resulting in optimal adiabatic evolution times for systems undergoing decoherence. We then show that fast control, even white noise or chaotic, can be used to mimic adiabaticity in a non-adiabatic regime. Since white noise and chaos occur naturally, no control is essentially required~\cite{JW}, in contrast to control approaches for adiabaticity relying on precisely engineered interventions, such as assistant Hamiltonians, dynamical decoupling~\cite{PhysRevLett.100.160506,PhysRevA.86.042333,Young:13}, or the quantum Zeno effect~\cite{PhysRevLett.108.080501}.

%%%%%%%%%%%%%%%
%\noindent \textbf{Results}
\section{Results}
%%%%%%%%%%%%%%%
In Sec.~\ref{subsec-A}-\ref{subsec-C} we present a derivation of an exact (approximation-free) TCL equation of motion for the projected eigenstate population. In Sec.~\ref{subsec-D} we invoke the adiabatic approximation and weak coupling in order to derive an appropriate perturbation theory. An example is presented and analyzed in detail in Sec.~\ref{example}. We show how our framework incorporates the closed system case in Sec.~\ref{closed}.

%%%%%%%%%%%%%%%%%%%%%%%%%%%%%%%%
\subsection{Open quantum systems in the adiabatic frame}
\label{subsec-A}
%%%%%%%%%%%%%%%%%%%%%%%%%%%%%%%%
Consider an $N$-level quantum system with a time-dependent Hamiltonian $H(t)$, with instantaneous eigenvalues $E_n(t)$ [$E_n(t) \leq E_{n+1}(t)$ $\forall n,t$] and eigenvectors $\ket{E_n(t)}$: $H(t) \ket{E_n(t)} = E_n(t) \ket{E_n(t)}$. In order to generalize the concept of adiabaticity to open quantum systems, it is convenient to adopt the superoperator formalism. We assume that the system is coupled to an environment and is described by a time-local master equation:
\begin{equation}
{\cal L} (t) |\rho (t) \rangle\rangle = {\partial_t} |\rho (t) \rangle\rangle .
\label{TLME}
\end{equation}
Here ${\cal L}(t) = -i {\cal H}(t)  + {\cal D}(t)$ is the Liouville superoperator, represented as an $N^2\times N^2$ matrix, and $| \rho (t) \rangle \rangle$ is the density operator associated with the system, represented as an $N^2\times 1$ vector (Hence it is represented by the double ket or bra notation. We reserve the ordinary ket or bra notation for the $N$-component vectors in Hilbert space). The superoperator ${\cal D}(t)$ denotes the contribution to ${\cal L}(t)$ arising from the coupling to the bath. Consider the basis of eigenvectors of the Hamiltonian superoperator ${\cal H}(t)$, defined through ${\cal H}(t) | \Phi_k (t) \rangle \rangle = \Lambda_k(t) |\Phi_k (t) \rangle \rangle$. The eigenvectors $|\Phi_k (t) \rangle \rangle$ of ${\cal H}(t)$ are the operators $\ket{E_n(t)}\langle E_m(t)|$, with eigenvalues $\Lambda_{k}=E_n(t)-E_m(t)$, where $k=m+nN$ and $m,n\in\{0,\dots,N-1\}$. The inner product of vectors $|u\rangle\rangle$ and $|v\rangle\rangle$ associated with operators $u$ and $v$, respectively, is defined as $\langle\langle u|v\rangle\rangle = \textrm{Tr}(u^\da v)$. The basis $\{|\Phi_k (t) \rangle \rangle \}$ of eigenstates of ${\cal H}(t)$ defines an ``adiabatic frame" in the open-system scenario. In this frame, the open system state can be expanded as
\begin{equation}
|\rho(t) \rangle \rangle = \sum_{k=0}^{N^2-1} r_k(t) e^{ -i \Theta_k(t) }  | \Phi_k (t) \rangle \rangle,
\label{rho-open}
\end{equation}
with $\Theta_k(t) = \int_0^t dt^\prime \Lambda_k(t^\prime) $ playing the role of a dynamical phase. Substituting the expansion~\eqref{rho-open} into the master equation~\eqref{TLME} yields a set of coupled differential equations for the coefficients $r_k(t)$, of the form
\beq
\partial_t \kett{R(t)} = {\cal L}^{(a)} \kett{R(t)} ,\quad {\cal L}^{(a)}(t) = -i {\cal H}^{(a)}(t) + {\cal D}^{(a)}(t),
\label{MEaf}
\eeq
with $\kett{R(t)} \equiv (r_0,\,r_1,\cdots,r_{N^2-1})^{T}$ (superscript $T$ denotes the transpose), where
\beq
{\cal H}_{kl}^{(a)}(t) = -i e^{ -i [\Theta_l(t)-\Theta_k(t)]}\braa{\Phi_k(t)}\partial_t\kett{\Phi_l(t)}
\label{H-ad-frame}
\eeq
are the matrix elements of the Hermitian matrix ${\cal H}^{(a)}(t)$ representing the (Hermitian) Hamiltonian superoperator in the adiabatic frame, and where
\begin{equation}
{\cal D}_{kl}^{(a)}(t) =  e^{ -i [\Theta_l(t)-\Theta_k(t)]}\braa{\Phi_k(t)}{\cal D}\kett{\Phi_l(t)} ,
\label{Lind-af2}
\end{equation}
are the matrix elements of the (generally non-Hermitian) matrix ${\cal D}^{(a)}(t)$ representing the decoherence superoperator in the adiabatic frame.

%%%%%%%%%%%%%%%%
\subsection{Feshbach P-Q partitioning}
\label{subsec-B}
%%%%%%%%%%%%%%%%
Bearing in mind that closed-system adiabaticity is associated with the decoupled evolution of the eigenstates $\ket{E_n(t)}$, our aim here will be to similarly consider the decoupled evolution of the instantaneous eigenstates of ${\cal H}(t)$ corresponding to the eigenprojections $\ketbra{E_n(t)}{E_n(t)}$ of $H(t)$. Specifically, denoting the target eigenstate of $H(t)$ by $|E_0(t)\rangle$, and assuming henceforth that it is non-degenerate, we will be interested in the decoupled evolution of the eigenprojection $\kett{\Phi_0 (t)} = \ketbra{E_0(t)}{E_0(t)}$, captured in the adiabatic frame by the population coefficient $r_0(t)$. We employ the Feshbach P-Q partitioning technique, introducing the projection operators $\map = 1\oplus \mathbb{0}_{N^2-1}$ and $\maq = \mai-\map = 0\oplus \mathbb{1}_{N^2-1}$ where $\mathbb{0}_{N^2-1}$ and $\mathbb{1}_{N^2-1}$ denote the $(N^2-1)\times(N^2-1)$ null and identity matrices, respectively, and ${\cal P}$ projects the system onto the target eigensubspace. We thus decompose both the adiabatic frame Hamiltonian ${\cal H}^{(a)}(t)$ and decoherence superoperator ${\cal D}^{(a)}(t)$ (dropping the explicit time-dependence) as
\bes
\begin{align}
\!\!{\cal H}^{(a)} & ={\cal H}_0+{\cal H}_1; \,\,
{\cal H}_0 =g_{\cal H}+e_{\cal H}, \,\, {\cal H}_1=W_{\cal H}+W^\dagger_{\cal H} \\
\!\!{\cal D}^{(a)} & ={\cal D}_0+{\cal D}_1;\,\,
{\cal D}_0=g_{\cal D}+e_{\cal D},\,\, {\cal D}_1=W_{\cal D}+V_{\cal D}\ ,
\end{align}
\ees
where $g$ and $e$ denote the target eigenstate (e.g., the ground state) and the remaining eigenstates (e.g., the excited states), respectively, and ${\cal H}_0,{\cal D}_0$ and ${\cal H}_1,{\cal D}_1$ denote block-diagonal and block-off-diagonal contributions, respectively, with
\bes
\label{geWD}
\begin{align}
g_{\cal H}&=\map {\cal H}^{(a)}\map\ , \,\, e_{\cal H}=\maq {\cal H}^{(a)}\maq\ , \,\,   \\
W_{\cal H}&=\maq {\cal H}^{(a)}\map\ , \,\, W^\dagger_{\cal H}=\map {\cal H}^{(a)}\maq; \label{geWDa} \\
g_{\cal D}&=\map {\cal D}^{(a)}\map\ , \,\,  e_{\cal D}=\maq {\cal D}^{(a)}\maq\ , \,\,    \\
W_{\cal D}&=\maq {\cal D}^{(a)}\map\ , \,\,  V_{\cal D}=\map{\cal D}^{(a)}\maq \ . \label{geWDb}
\end{align}
\ees
Note that in general $V_{\cal D} \neq W^\dagger_{\cal D}$.

%%%%%%%%%%%%%%%%%%%%%%%%%%%%%%
\subsection{Time-convolutionless dynamics for eigenstate tracking}
\label{subsec-C}
%%%%%%%%%%%%%%%%%%%%%%%%%%%%%%
We now derive an exact, time-convolutionless (TCL) dynamical equation for the target eigenstate population $r_0(t)$, using a method inspired by the approach in Ref.~\cite{Breuer:book}. Let ${\cal U}_0(t)$ denote the evolution operator associated with ${\cal H}_0(t)$, i.e.,
\begin{equation}
\label{U0GgGe}
{\cal U}_0(t)= {\cal G}_{g}(t,0) + {\cal G}_e(t,0)\ ,
\end{equation}
where
\begin{equation}
\label{GgGe}
{\cal G}_{g}(t,t')\equiv e^{-i\int_{t'}^t  g_{\cal H}(s) ds}, \,\,
{\cal G}_e(t,t')\equiv\mathcal{T}[e^{-i\int_{t'}^t  e_{\cal H} (s) ds}]\ ,
\end{equation}
with $\mathcal{T}$ denoting forward time-ordering. By working in the interaction picture with respect to the block-diagonal Hamiltonian part ${\cal H}_0$, Eq.~\eqref{MEaf} becomes
\begin{equation}
\label{MEafIP}
\partial_t\kett{\chi(t)}={\cal L}_I(t) \kett{\chi(t)}\ ,
\end{equation}
where $\kett{\chi(t)}={\cal U}^\dagger_0(t)\kett{R(t)}$, and
\begin{equation}
{\cal L}_I(t) = -i {\cal H}_I(t) + {\cal D}_I(t), \quad
\end{equation}
with
\begin{equation}
\label{eq:I-pic}
{\cal H}_I(t)= {\cal U}^\dagger_0(t) {\cal H}_1(t){\cal U}_0(t) , \quad
{\cal D}_I(t)= {\cal U}^\dagger_0(t) {\cal D}^{(a)}(t) {\cal U}_0(t) \ .
\end{equation}
Note that $\mc{H}_I(t)$ is purely block off-diagonal, i.e., $\map\mc{H}_I(t)\map = \maq\mc{H}_I(t)\maq = 0$. By projecting Eq.~\eqref{MEafIP} over $\map$ and $\maq$ [i.e., inserting $\map+\maq=\openone$ into Eq.~\eqref{MEafIP}], we decompose the time-local master equation into ``relevant" (target eigenstate) and ``irrelevant" (the remaining eigenstates) components, respectively:
\bes
\begin{align}
\label{Opphi}
\pa_t\map\kett{\chi(t)}&= \map {\cal L}_I(t)  \maq \kett{\chi(t)} + \map {\cal L}_I(t) \map \kett{\chi(t)},\\
\label{Oqqphi}
\pa_t\maq\kett{\chi(t)}&= \maq {\cal L}_I(t)  \map \kett{\chi(t)} + \maq {\cal L}_I(t) \maq \kett{\chi(t)} .
\end{align}
\ees
Next, we introduce the propagator
\beq
\mathcal{G}(t,t^\prime)\equiv\mathcal{T}\left\{\exp\left[\int_{t^\prime}^t\maq {\cal L}_I(x)dx\right]\right\}\ .
\eeq
We show in Appendix~\ref{sec:check-sol}  that the formal solution to Eq.~\eqref{Oqqphi} is
\beq
\maq \kett{\chi(t)} \hspace{-0.1cm}  = \mc{G}(t,0) \maq \kett{\chi(0)} + \int_0^t \hspace{-0.2cm}  \mc{G}(t,t') \mc{Q} \mc{L}_I(t') \mc{P}\kett{\chi(t')} dt' .
\label{Oqphi}
\eeq
The first term on the r.h.s. of Eq.~\eqref{Oqphi} vanishes if, as we assume from now on, that the system is prepared in the initial eigenstate $\kett{\Phi_0(0)}$ of ${\cal H}(0)$, so that $\maq\kett{\chi(0)}=0$.

Inserting Eq.~\eqref{Oqphi} into Eq.~\eqref{Opphi} we obtain the Nakajima-Zwanzig equation for the target eigenstate component:
\begin{eqnarray}
\pa_t\map\kett{\chi(t)}&=&  \int_0^t \map {\cal L}_I(t) \mc{G}(t,t') \mc{Q} \mc{L}_I(t') \mc{P}\kett{\chi(t')} dt'  \nonumber \\
&+& \map {\cal L}_I(t) \map \kett{\chi(t)} \ .
\label{eq:NZ}
\end{eqnarray}
This result is remarkable, since it gives an exact representation of the ground state evolution of an open quantum system. However, it involves solving a rather complicated integro-differential equation.

To make further progress, in particular to obtain a time-local dynamical equation, we define
\begin{equation}
\kett{\chi(t^\prime)}={\cal V}_I^{-1}(t,t^\prime)\kett{\chi(t)}, \quad
\end{equation}
where
\begin{equation}
{\cal V}_I^{-1}(t,t^\prime)=\mathcal{T}_{\leftarrow}\left\{\exp\left[- \int_{t^\prime}^t   {\cal L}_I(x) dx\right]\right\} ,
\end{equation}
with $\mathcal{T}_{\leftarrow}$ denoting reverse time-ordering. This allows us to rewrite Eq.~\eqref{Oqphi} as $\maq \kett{\chi(t)} = \Sigma(t) (\map+\maq)\kett{\chi(t)}$,
where
\beq
\Sigma(t) \equiv \int_0^t \mc{G}(t,t') \mc{Q} \mc{L}_I(t') \mc{P}{\cal V}_I^{-1}(t,t^\prime) dt' .
\label{eq:Sigma}
\eeq
Thus
\begin{equation}
\maq \kett{\chi(t)} = \left[ \openone - \Sigma(t) \right]^{-1}  \Sigma(t)  \map \kett{\chi(t)} ,
\label{OTCLQ}
\end{equation}
As discussed in Appendix~\ref{sec:ad-pert}, the invertibility of $\left[ \openone - \Sigma(t) \right]$ is ensured in the closed system case due to the fact that it can be treated as a perturbation of the identity in the large $T$ limit, and in the open system case if in addition the system-bath interaction is weak. Substituting Eq.~\eqref{OTCLQ} into Eq.~\eqref{Opphi}, we obtain
$\pa_t\map\kett{\chi(t)}= \map \mc{L}_I(t)\left[\left(\openone - \Sigma(t)\right)^{-1}\Sigma(t) + \openone\right] \map \kett{\chi(t)}$, which simplifies to
\bes
\label{eq:TCL-result}
\begin{align}
\label{OKtcl}
\pa_t\map\kett{\chi(t)} &=\mathcal{K}(t)\map\kett{\chi(t)} , \\
\mathcal{K}(t) &= \map {\cal L}_I(t)\left[\openone - \Sigma(t)\right]^{-1}  \map .
\label{OKgen}
\end{align}
\ees
Here $\mc{K}(t)$ is the TCL generator. Equation~\eqref{eq:TCL-result} constitutes our main result: \emph{an exact, time-convolutionless dynamical equation for the (projected) target eigenstate population.} This time-local result is clearly a significant simplification compared to the Nakajima-Zwanzig equation [Eq.~\eqref{eq:NZ}], but it is still difficult, in general, to calculate the TCL generator. To make further progress we next pursue a perturbative approach. Specifically, we shall consider an adiabatic (long time) approximation along with weak coupling between the system and the bath.

%%%%%%%%%%%%%%%%%%%%%%%%%%%
\subsection{Weak coupling and adiabatic dynamics}
\label{subsec-D}
%%%%%%%%%%%%%%%%%%%%%%%%%%%
As a first step towards a perturbative expansion we write $ \left[\openone - \Sigma(t)  \right]^{-1}$ as a geometric series. Using Eq.~\eqref{OKgen}, this yields
\begin{equation}
\mathcal{K}(t) = \sum_{n=0}^{\infty} {\cal P} {\cal L}_I (t) \left[ \Sigma(t)\right]^{n+1} {\cal P} +  \map {\cal D}_I(t) \map\ ,
\label{TCLs}
\end{equation}
where we also used the fact that $\map {\cal H}_I(t) \map = 0$. We now assume that the contribution of the decoherence superoperator $\mc{D}_I$ is perturbative due to weak system-bath coupling. We can then use Eq.~\eqref{eq:Sigma} to expand $\Sigma(t)$ in powers of $\mc{L}_I(t)$.  Since $\mc{G}(t,t')= \openone + \mc{O}(\mc{Q}\mc{L}_I)$ and ${\cal V}_I^{-1}(t,t^\prime) = \openone + \mc{O}(\mc{L}_I)$, the lowest order term in $\mc{L}_I(t)$ for $\Sigma(t)$ is $\Sigma^{(1)}(t)= \int_0^t dt'{\cal Q} {\cal L}_I(t') {\cal P}$.

From this point on, it is useful to split up $\mc{L}_I(t)$ into the Hamiltonian and decoherence superoperators ${\cal H}_I(t)$ and $\mc{D}_I(t)$, respectively [Eq.~\eqref{eq:I-pic}]. Moreover, let $s=t/T\in[0,1]$ denote the normalized time, with $T$ the total evolution time, and let us replace $t$ by the pair $(s,T)$ in order to prepare for an expansion in $1/T$ (adiabatic perturbation theory~\cite{Teufel:book}). We will proceed by keeping contributions up to leading order in $1/T$, which will provide a reliable approximation for large $T$. As shown below [Eq.~\eqref{21a}], this corresponds to keeping terms linear in ${\cal H}_I(s',T)$.

By inserting $\Sigma^{(1)}(s,T)$ into Eq.~\eqref{TCLs} and simplifying using $\mc{Q}=\openone - \map$, we find a term that represents the zeroth-order decoherence contribution to the TCL generator (i.e., that does not depend on $\mc{D}_I$ at all and hence describes the system in the absence of the bath),
\begin{equation}
\mathcal{K}^{(0)}(s,T) = -  T \int_0^s ds' \map {\cal H}_I(s,T) {\cal H}_I(s',T) \map \ ,
\label{eq:K0}
\end{equation}
while the first-order decoherence contribution (linear in $\mc{D}_I$) takes the form
\begin{align}
\label{eq:K1}
&\mathcal{K}^{(1)}(s,T) = \map {\cal D}_I(s,T) \map \\
&- i T \hspace{-0.1cm} \int_0^s \hspace{-0.2cm} ds' \map \left[ {\cal D}_I(s,T) {\cal H}_I(s',T) + {\cal H}_I(s,T) {\cal D}_I(s',T) \right] \map . \nonumber
\end{align}
Quadratic and higher order terms in $\mc{D}_I$ can easily be written down by following the same strategy.

\begin{figure}[htbp]
\centering
\subfigure[]{\includegraphics[width=3in]{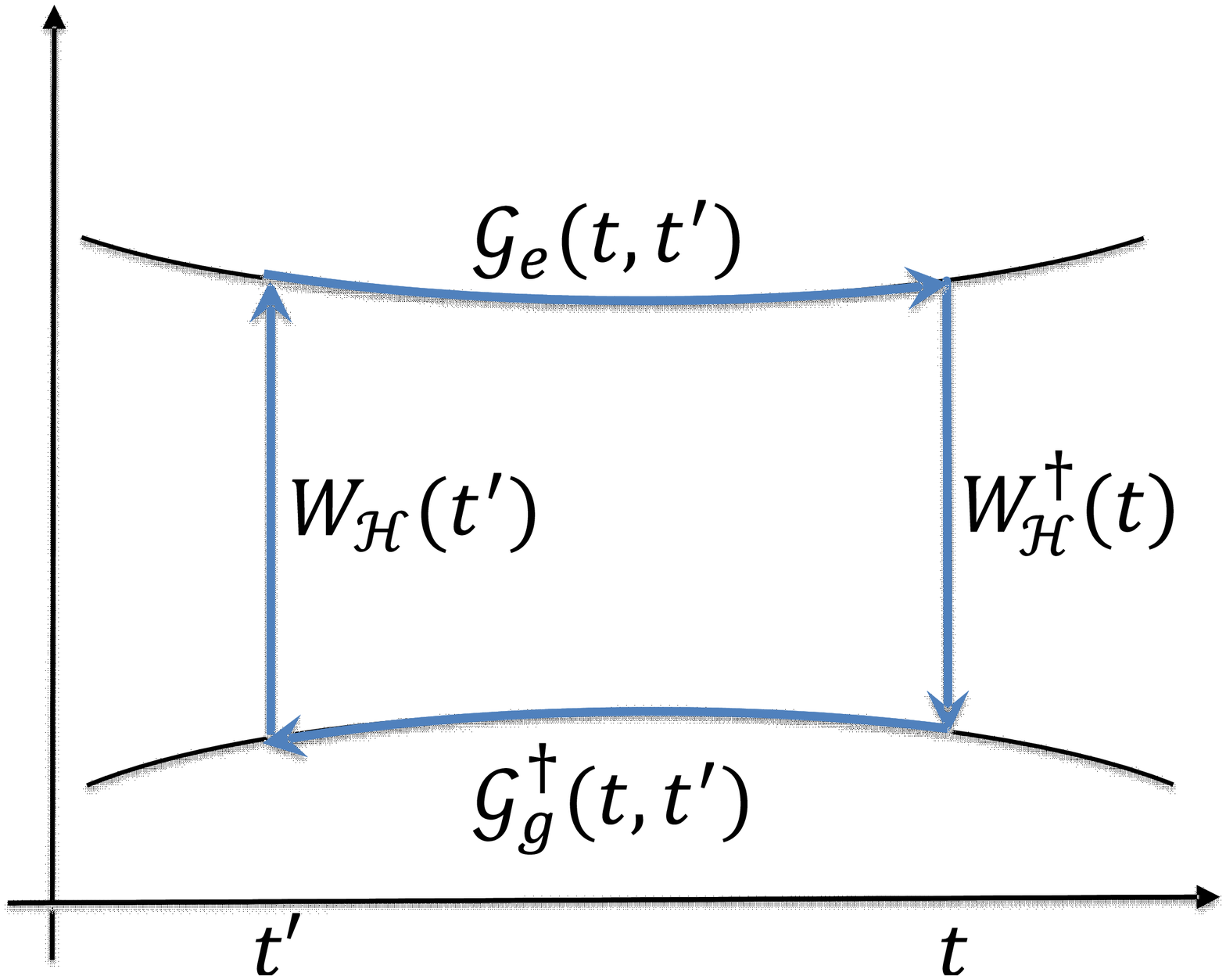}}
\subfigure[]{\includegraphics[width=3in]{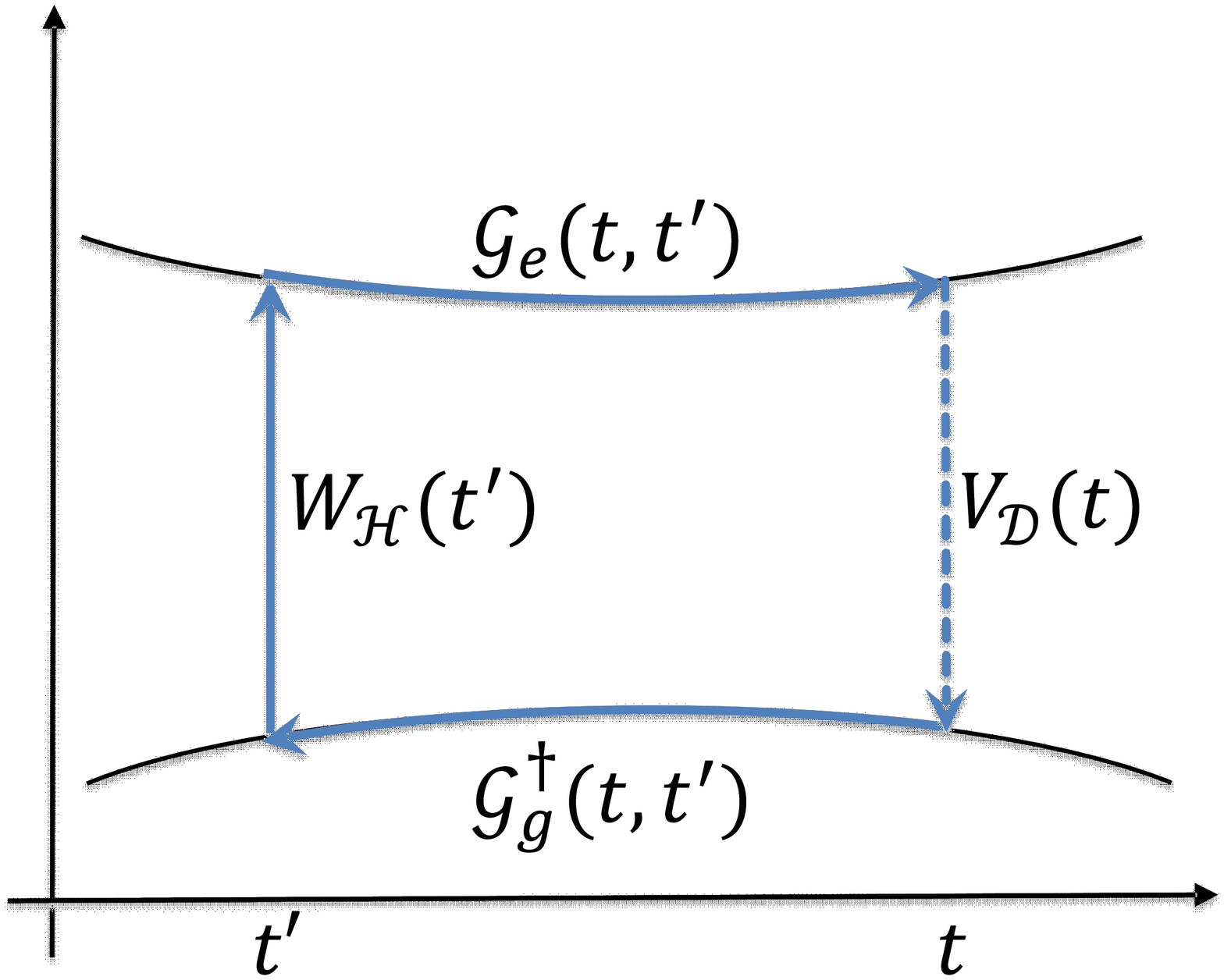}}
\subfigure[]{\includegraphics[width=3in]{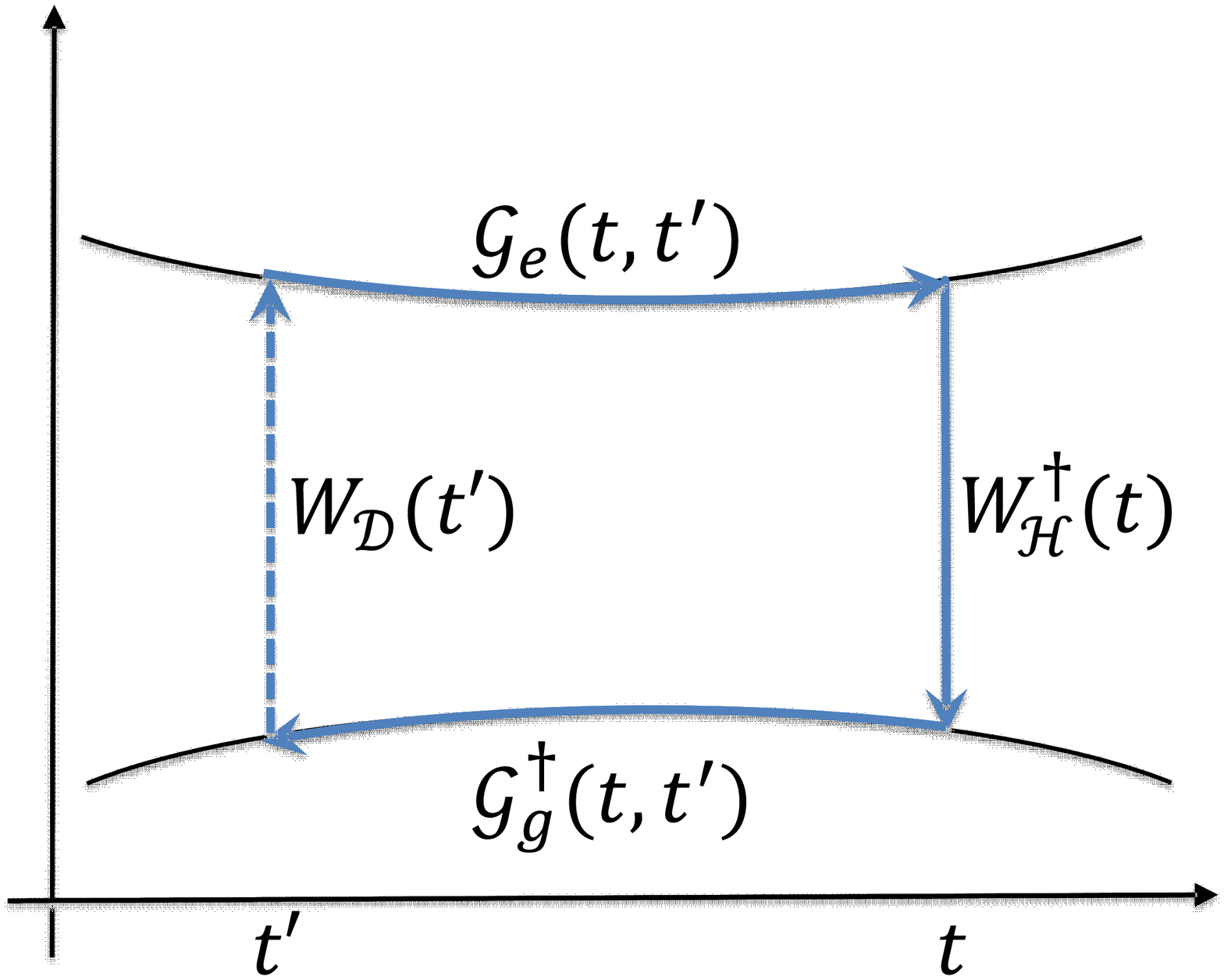}}
%\subfigure[]{\includegraphics[width=0.32\textwidth]{diagram1.pdf}\label{fig:diagram1}}\hspace{-1cm}
%\subfigure[]{\includegraphics[width=0.32\textwidth]{diagram2.pdf}\label{fig:diagram2}}\hspace{-1cm}
%\subfigure[]{\includegraphics[width=0.32\textwidth]{diagram3.pdf}\label{fig:diagram3}}
\caption{Diagrams illustrating the transitions described by Eq.~\eqref{eq:gf-open}. (a) Purely unitary evolution, involving an excitation from the target state to the remaining eigenstate manifold, and the reverse process. (b) Open system evolution involving non-unitary decay to the target eigenstate along with unitary evolution along the other three paths. (c) Open system evolution involving non-unitary excitation from the target eigenstate along with unitary evolution along the other three paths. These are the three lowest order processes in our perturbation theory.}
\label{fig:diagram}
\end{figure}

Let us now demonstrate that the standard adiabatic theorem for closed system is captured by the $\mathcal{K}^{(0)}$ term, while the competition between adiabaticity and decoherence is captured by $\mathcal{K}^{(1)}$. Indeed, using Eqs.~\eqref{eq:K0} and \eqref{eq:K1}, we find that to first order in $\mc{D}_I$, Eq.~\eqref{eq:TCL-result} reads
\bes
\label{Opphi3}
\begin{align}
\label{21a}
&\pa_s\map\kett{\chi(s)} = \nonumber \\
&- T^2  \map\int_0^s ds^\prime  {\cal H}_I(s,T) {\cal H}_I(s^\prime,T)\map\kett{\chi(s)}  \\
&+\map\left\{ T\mc{D}_I(s,T)-iT^2  \int_0^s ds^\prime \left[ {\cal D}_I(s,T) {\cal H}_I(s^\prime,T) \right. \right.\nonumber \\
&\left.\left.+ {\cal H}_I(s,T) {\cal D}_I(s^\prime,T)\right] \frac{}{}\right\} \map\kett{\chi(s)}\ .
\label{21b}
\end{align}
\ees
Note that due to the $\partial_t$ in Eq.~\eqref{H-ad-frame} each ${\cal H}_I$ term contributes a factor of $1/T$ in Eq.~\eqref{Opphi3}. Moreover, as we show in Appendix~\ref{sec:ad-pert}, after integration by parts both ${\cal H}_I(s',T)$ and ${\cal D}_I(s',T)$ can be written as a series whose lowest order is $\mc{O}(1/T)$. This is a consequence of the fact that the dynamical phase term in Eqs.~\eqref{H-ad-frame} and \eqref{Lind-af2} depends on $T$ as well: $\Theta_k(s,T) = T \int_0^s ds^\prime \Lambda_k(s^\prime,T)$. Therefore, the zeroth-order decoherence contribution [Eq.~\eqref{21a}] vanishes as $T^2/T^3 = 1/T$. This vanishing of the closed system component in the large $T$ limit is, of course, in agreement with the standard adiabatic theorem for closed systems. However, the same scaling arguments imply that the first-order decoherence contribution [Eq.~\eqref{21b}] survives and grows as $\mc{O}(T)$ for large $T$. This survival of a term that depletes the target eigenstate population even in the adiabatic (large $T$) limit is a purely open-system effect. It implies a competition between the adiabatic and  decoherence time-scales, resulting in an optimal finite time for the approximately adiabatic (decoupled) evolution of the system. This conclusion was first proposed using the Jordan block decoupling criterion in Ref.~\cite{PhysRevA.71.012331}, but here we see that it holds for the Hamiltonian eigenstate population. Experimental evidence for an optimal adiabatic time was reported in Ref.~\cite{Steffen:2003ys}.

We now rewrite Eq.~\eqref{Opphi3} in terms of the original time variable $t$ and rotate it back to the Schr\"odinger picture. Using $\kett{\chi(t)}={\cal U}^\dagger_0(t)\kett{R(t)}$, we have $\pa_t\map\kett{\chi(t)} =  {\cal G}_{g}^\da(t,0) \left[ \pa_t \map | R(t)\ra\ra + i g_{\cal H}(t) \map  | R(t)\ra\ra  \right]$. After some algebra using the definitions of ${\cal H}_I(t)$ and ${\cal D}_I(t)$ [Eq.~\eqref{eq:I-pic}], the zeroth [Eq.~\eqref{eq:K0}] and first order [Eq.~\eqref{eq:K1}] perturbative decoherence contributions can be written as $\mathcal{K}^{(0)}(t) = - {\cal G}_{g}^\da(t,0) \int_0^tdt^\prime h(t,t')$ and $\mathcal{K}^{(1)}(t) = - {\cal G}_{g}^\da(t,0) \int_0^tdt^\prime f(t,t^\prime)+{\cal G}_{g}^\da(t,0)g_{\cal D}(t)$, with
\bes
\label{eq:gf-open}
\begin{align}
h(t,t') &= {W}^\dagger_{\cal H}(t) {\cal G}_e(t,t^\prime) {W}_{\cal H}(t^\prime) {\cal G}_{g}^\da(t,t^\prime)
\label{gopen}\\
f(t,t^\prime) &= i \left[ {W}^\dagger_{\cal H}(t) {\cal G}_e(t,t^\prime) {W}_{\cal D}(t^\prime) \right. \nonumber \\
&\left.\,\,\,\,\,\,\,\,\,\,+ V_{\cal D}(t) {\cal G}_e(t,t^\prime) {W}_{\cal H}(t^\prime)\right] {\cal G}_{g}^\da(t,t^\prime) .
\label{fopen}
\end{align}
\ees
This leads to our second main result: a time-local master equation in the Schr\"odinger picture for the target eigenstate population. Namely, Eq.~\eqref{Opphi3} can now be rewritten as
\begin{align}
\pa_t \map | R(t)\ra\ra =&-\map \left[ \left(i g_{\cal H}(t) +\hspace{-0.1cm} \int_0^t \hspace{-0.2cm} dt^\prime h(t,t')\right) \right. \nonumber \\
&\left.- \left( g_{\cal D}(t) - \hspace{-0.1cm} \int_0^t \hspace{-0.2cm} dt^\prime f(t,t^\prime)\right) \right] \map | R(t)\ra\ra,
\label{de-open}
\end{align}
with $\map \kett{R(t)} \equiv (r_0(t),\,0,\cdots,0)^{T}$ being the target eigenstate population, and we have separately grouped the contributions due to the Hamiltonian and decohering parts. Here [recall Eq.~\eqref{geWD}]
\bes
\begin{align}
g_{\cal H}(t) &= {\cal H}_{00}^{(a)}(t) = -i \braa{\Phi_0(t)}\partial_t\kett{\Phi_0(t)}, \\
g_{\cal D}(t) &= \braa{\Phi_0(t)}\mc{D}^{(a)}\kett{\Phi_0(t)}
\end{align}
\ees
are, respectively, the target eigenstate matrix elements of the adiabatic frame Hamiltonian and decoherence superoperators; the terms $h(t,t')$ and $f(t,t^\prime)$ are associated with the $\mc{H}^{(a)}$-dependent and $\mc{D}^{(a)}$-dependent level couplings in the dynamical evolution, respectively.

Note that $h(t,t')$ is responsible for non-adiabatic closed system dynamics: as is evident from Eq.~\eqref{gopen}, it describes a unitary evolution in the target eigenstate [${\cal G}_{g}^\da(t,t^\prime)$], followed by a transition to the remaining eigenstate manifold [${W}_{\cal H}(t^\prime)$], unitary evolution in that manifold [${\cal G}_e(t,t^\prime)$], and finally a transition back down to the target eigenstate [${W}^\dagger_{\cal H}(t)$; see Fig.~\ref{fig:diagram}(a)]. Similarly, $\int_0^t dt^\prime f(t,t^\prime)$ is responsible for non-adiabatic open system dynamics; Eq.~\eqref{fopen} shows that this contribution is mediated by a transition back to the target eigenstate [${V}_{\cal D}(t')$; see Fig.~\ref{fig:diagram}(b)] or out of the target eigenstate [${W}_{\cal D}(t)$; see Fig.~\ref{fig:diagram}(c)], both generated by the decoherence operator. It is clear that in higher order perturbation theory each term will contain several such transitions, including a mixing of transitions generated by the Hamiltonian and decoherence superoperators. An evolution that perfectly preserves the target eigenstate (at this level of perturbation theory) would thus require the vanishing of both $\int_0^tdt^\prime h(t,t')$ and $\int_0^tdt^\prime f(t,t')$. It follows from our earlier arguments that $\int_0^tdt^\prime h(t,t')$ vanishes for large $T$ (adiabatic evolution in the closed system limit), while the decoherence contribution $\int_0^t dt^\prime f(t,t^\prime)$ need not. However, as we shall see in examples below, adiabatic evolution can be mimicked by introducing appropriate fast modulations that cause both integrals to vanish, without the need for counter-diabatic driving.

\section{Applications}

%%%%%%%%%%%%%%%%%%%%%%%%%%%%%%%%%%
\subsection{Open-system quasi-adiabatic evolution of a qubit}
\label{example}
%%%%%%%%%%%%%%%%%%%%%%%%%%%%%%%%%%

To illustrate the general theory we have developed for adiabaticity in open systems, we turn now to the consideration of the decay of a single qubit. In this section we demonstrate control protocols that allow for tracking of an excited state. Moreover, we show that these protocols are insensitive to implementation details. Specifically, we assume that the time-dependent system Hamiltonian is
\beq
H(t)=J(t)\left[\cos\left(\frac{\pi}{2T}t\right)\si_z+\sin\left(\frac{\pi}{2T}t\right)\si_x\right] \ ,
\label{eq:H(t)}
\eeq
whose eigenvectors and eigenvalues are $\ket{E_+(t)}=\cos\left(\frac{\pi}{4T}t\right)\ket{0}-\sin\left(\frac{\pi}{4T}t\right)\ket{1}$, $\ket{E_-(t)}=\sin\left(\frac{\pi}{4T}t\right)\ket{0}+\cos\left(\frac{\pi}{4T}t\right)\ket{1}$ and $E_{\pm}=\mp J(t)$, respectively, with $\sigma_z |0\rangle=-|0\rangle$ and $\sigma_z |1\rangle=|1\rangle$. Thus the time-dependent gap is $2J(t)$, where $J(t)$ is assumed to be a controllable parameter. We transform to the rotating frame provided by the eigenstate basis $\{\ket{E_\pm(t)}\}$ and assume a spin-boson model with system-bath interaction $H_{SB}(t)$ and bath Hamiltonian $H_B$:
\bes
\begin{align}
H_{SB}(t) &= \sum_k \left(g_k\si_+(t)b_k+g_k^*\si_-(t)b_k^\da \right), \\
H_{B} &= \sum_k\om_kb_k^\da b_k,
\end{align}
\ees
where the creation and annihilation operators $b_k$ and $b_k^\da$ satisfy bosonic commutation relations $[b_k, b_{k'}^\da]=\de_{kk'}$ and the coupling operators are $\si_+(t)\equiv|E_-(t)\ra\la E_+(t)|$ and $\si_-(t)\equiv \si_+^\dagger(t)$.
The evolution of the qubit is then described by the exact master equation~\cite{Vacchini:2010fv}
\begin{eqnarray}\label{exactME}
\pa_t\rho(t)&=&-i[J(t)+{\cal S}(t)][\sigma_z(t), \rho(t)] \nonumber \\
&&\hspace{-1.6cm}+\ka(t) \left[\sigma_-(t)\rho(t)\si_+(t)-\frac{1}{2} \left\{\sigma_+(t)\sigma_-(t),\rho(t)\right\}\right],
\end{eqnarray}
where $\si_z(t)\equiv |E_-(t)\ra\la E_-(t)|-|E_+(t)\ra\la E_+(t)|$, $\ka(t)\equiv-2{\rm Re}[\dot{c}_+/c_+]$, and ${\cal S}(t)\equiv-{\rm Im}[\dot{c}_+/c_+]$ (dot denotes time derivative). Here $c_+(t)\equiv \ti{c}_+(t)e^{i\int_0^tdsJ(s)}$, where $\ti{c}_+(t)$ is the solution of $\dot{\ti{c}}_+(t)+\int_0^tds\ti{\al}(t-s)\ti{c}_+(s)=0$, and $\ti{\al}(t-s)\equiv\al(t-s)e^{-2i\int_s^tdxJ(x)}$, with $\al(t-s)\equiv\sum_k|g_k|^2e^{-i\om_k(t-s)}$ denoting the bath correlation function.

The system is prepared in the initial excited state $|E_-(0)\rangle=|1\rangle$. We use the Uhlmann fidelity ${\cal F}\equiv\sqrt{\la E_-(T)|\rho(T)|E_-(T)\ra}=\sqrt{|r_0(T)|}$ as a measure of adiabaticity, where $r_0(T)$ is the excited state population at the final time $T$. Let us denote the average gap between the ground and excited states by
\begin{equation}
\ti{J}(t)\equiv\frac{1}{t}\int_0^tdt' J(t').
\end{equation}
By computing $h(t, t')$ and $f(t,t')$  to leading order using Eqs.~\eqref{gopen} and \eqref{fopen}, respectively, we obtain (see Appendix~\ref{sec:check-eqhf} for details), after approximating $\ti{J}$ by a constant function,
\bes
\label{eq:hf}
\begin{align}
\left. h(t,t')\right|_{11}&\approx  \frac{\pi^2}{8T^2}\cos[2(t-t')\ti{J}]\ ,   \\
\left. f(t,t')\right|_{11} &\approx  -\frac{\pi^2\ka(t')}{16T^2\ti{J}}\sin[2(t-t')\ti{J}]\ ,
\end{align}
\ees
where the scalars $\left. h(t,t')\right|_{11}$ and $\left. f(t,t')\right|_{11}$ are the nonvanishing $(1,1)$ matrix elements of the projected matrices $[\map \, h(t,t')\, \map]$ and $[\map\, f(t,t')\, \map]$, respectively. Note that both the closed-system contribution $h(t,t')$ and the open-system contribution $f(t,t')$ decay as $1/T^2$ to leading order. This is because the order $1/T$ contribution in $f(t,t')$ vanishes for this example (as shown in  Appendix~\ref{sec:check-eqhf}). Another important observation that follows from Eq.~\eqref{eq:hf} is that a sufficiently large $\ti{J}$ implies the vanishing of the integrals [Eq.~\eqref{de-open}] of both $h(t,t')$ and $f(t,t')$ due to the highly oscillatory nature of the integrand (i.e., the Riemann-Lebesgue lemma~\cite{Ryzhik}). Recall that these terms are responsible for transitions out of the eigenstate considered (Fig.~\ref{fig:diagram}). This illustrates that \emph{adiabaticity may be enforced via active control}, as shown in the closed system case in Ref.~\cite{Jing:2014uq}. Our results show that this conclusion persists even in the open system case (see also Refs.~\cite{PhysRevLett.100.160506,PhysRevA.86.042333,Young:13}).

Next, we compute the exact fidelity ${\cal F}$ by solving the master equation~\eqref{exactME}, and the TCL-approximation ${\cal F}_{\rm TCL}$ for the fidelity by substituting Eq.~\eqref{eq:hf} into Eq.~\eqref{de-open}. The results are, respectively (see Appendix~\ref{sec:check-eqhf} for details)
\bes\label{Fiall}
\begin{align}\label{Fi}
{\cal F}&=\exp\left[-\frac{1}{2}\int_0^Tdt\ \ka(t)\right]\ , \\
\label{FiTCL}
{\cal F}_{\rm TCL}&\approx \exp\left[-\frac{1}{2}\int_0^Tdt\ \ka(t)+\left(\frac{\pi}{8}\right)^2\frac{\left(\cos2\ti{J}T-1\right)}{(\ti{J}T)^2}\right]  \ .
\end{align}
\ees
As a concrete application, we now assume an environment with correlation function $\alpha (t,s) = \frac{\Gamma \gamma}{2} e^{-\gamma |t-s|}$. where $\ga$ parameterizes the memory of the environment ($1/\ga$ is proportional to the memory time) and $\Ga$ quantifies the system-bath coupling strength. We consider a fixed $\Ga$ and use $\Ga t$ to represent a dimensionless time variable. The control function $J(t)$ is taken as
\begin{equation}
J(t)=J_0+\Omega(t),
\end{equation}
which will be associated with two distinct control procedures: a periodic pulse sequence and biased Poissonian continuous-time white noise~\cite{JW,Kim:2007hb,Spiechowicz:2013yf}. Specifically, we consider the following scenarios: (a) a periodic rectangular pulse sequence $\Omega(t)=\Psi/\De$ for $n\chi-\De<\Ga t<n\chi$ and $\Omega(t)=0$ elsewhere, where $n \geq 1$ is an integer, $\Psi$ denotes the pulse amplitude, and $\De$ is the duration of the pulse in one period $\chi$; (b) white noise $\Omega(t)=\sum_{j=1}^K \Omega_j\de(t-t_j)$, where the times $t_j$ and the amplitudes $\Omega_j$ are random during the duration $\De$ of the pulse and the latter vanish afterwards, i.e., they vanish in the dark time $[n\chi,(n+1)\chi-\De]$. The amplitudes satisfy $M[\Om_j]=\Ga$ for each duration $\De$, with $M[\cdot]$ denoting an ensemble average. Note that the control merely rescales the eigenvalues $E_\pm$ of $H(t)$ [Eq.~\eqref{eq:H(t)}], but it does not modify the instantaneous eigenstates. It can be realized, e.g., for spin systems, by changing the splitting of the system energy levels via an external magnetic field.

\begin{figure}[htbp]
\centering
\includegraphics[width=3.4in]{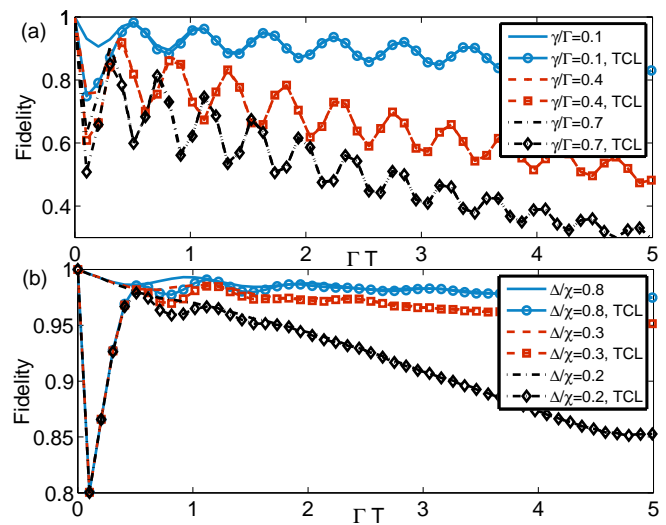}
\caption{Excited state fidelity for a qubit evolving quasi-adiabatically in the presence of a dissipative environment as a function of the dimensionless time $\Gamma T$, for the model described in the text. (a) Periodic rectangular pulse sequence for several values of $\gamma$ (the inverse bath memory time), with $\Delta /\chi = 0.4$; (b) White noise pulse sequence for several ratios $\Delta/\chi$ of pulse duration to sequence period, with $\ga=0.5\Ga$. The control parameters are given by $J_0=\Psi=\Ga $ and $\chi = 0.02\Ga t$, with time slices chosen such that sequence length $K \in [6,16]$. Convergence is obtained by averaging over 200 realizations of the white noise sequence. Curves without symbols are the exact fidelity results [Eq.~\eqref{Fi}], while curves with symbols are the perturbative TCL results [Eq.~\eqref{FiTCL}].}
\label{fig:fopen}
\end{figure}

We illustrate our results by comparing the exact and TCL-approximation cases in Fig.~\ref{fig:fopen}, where we plot the excited state fidelity for the model specified above. Fig.~\ref{fig:fopen}(a), for periodic control, shows that the longer the bath memory time $1/\ga$ is, the slower the overall fidelity decays. The fidelity exhibits oscillations, showing that it can be optimized locally in time. This can be achieved by using a protocol reminiscent of previous work on the use of dynamical decoupling to enhance adiabaticity in open quantum systems~\cite{PhysRevLett.100.160506,PhysRevA.86.042333,Young:13}, but without requiring any encoding. The symbol-free and symbol-marked curves denote the exact and TCL-approximation fidelities, respectively. It can be seen that, while the TCL approximation overestimates the decay rate of the fidelity for small $\Ga T$, the difference between these fidelities quickly tends to vanish with larger $\Ga T$, especially for small $\ga/\Ga$ values. This is consistent with Eq.~\eqref{FiTCL}, which shows that the two fidelities converge to the same value for large $\Ga T$. In this regime, the perturbative TCL method provides a rather accurate description of the fidelity, irrespective of the choice of $\ga/\Ga$. Figure~\ref{fig:fopen}(b) shows the fidelity under white noise control, for different ratios of the pulse duration time and the period. Qualitatively, the results are similar to those in the case of periodic control, i.e., the exact and the approximated fidelities tend to rapidly converge in the regime of large $\Ga T$. However, the oscillations seen in the periodic control case are absent, and the fidelity tends to monotonically decrease for sufficiently large $\Ga T$. This suggests that, in the presence of white noise control, it is harder to find optimized fidelity and evolution time pairs. The fidelity improves monotonically in terms of the ratio of the pulse duration to pulse sequence period, meaning that more control (i.e., more random $\delta$ functions) improves the fidelity, despite the control being stochastic.

%%%%%%%%%%%%%%%%%%%%%%%%%%%%%%%%%%%%
\subsection{The closed system limit and shortcuts to adiabaticity}
\label{closed}
%%%%%%%%%%%%%%%%%%%%%%%%%%%%%%%%%%%%
%\textbf{Transitionless evolution in a closed system.}

The formalism we have developed can also be applied in the closed system limit, which can be obtained by simply requiring that $\left( g_{\cal D}(t) - \int_0^t dt^\prime f(t,t^\prime)\right)  \mapsto \mathbb{0}_{N^2}$ in Eq.~\eqref{de-open}. However, in this scenario, it is more convenient to abandon the superoperator formalism. To set up the conventional Hilbert space notation, we expand the state vector in the adiabatic frame as $|\psi(t)\rangle = \sum_{n=0}^{N-1} c_n(t) e^{-i\theta_n(t)}|E_n(t)\ra$, where $\theta_n(t)\equiv \int_0^tdt'\ E_n(t')$ is the dynamical phase. Then, the Schr\"odinger equation $i \partial_t |\psi(t)\ra = H(t) |\psi(t)\ra$ ($\hbar\equiv1$) yields~\cite{Jing:2014uq}
\bes
\label{cm}
\begin{align}
\pa_t c_m(t) &= -i \sum_{n=0}^{N-1} H^{(a)}_{mn}(t) c_n(t), \\
 H^{(a)}_{mn}(t)&\equiv -i e^{-i(\theta_n-\theta_m)} \la E_m|\partial_t|E_n\ra
\end{align}
\ees
where the Hermitian $N\times N$ matrix $H^{(a)}(t)$ is the Hamiltonian in the adiabatic frame~\cite{Wuzy,tong_quantitative_2005,Tong:07}. Equation~\eqref{cm} thus represents a set of differential equations for the components of the vector $|C(t)\ra\equiv[c_0,\,c_2,\cdots,c_{N-1}]^{T}$. Assuming non-degeneracy, the target state amplitude can be taken as that of the ground state, $c_0(t)$. We may now repeat the earlier derivation (with the projection operators $\map = 1\oplus \mathbb{0}_{N-1}$ and $\maq = \mai-\map = 0\oplus \mathbb{1}_{N-1}$), or skip directly to the leading order of the closed system version of the TCL dynamical equation [Eq.~\eqref{de-open}]:
\begin{equation}
\label{TCLE}
\pa_t \map |C(t)\ra=-\map \left( i g_{H}(t)+\int_0^t dt^\prime h(t,t')\right) \map |C(t)\ra\ ,
\end{equation}
where $h(t,t')$ is the closed-system transition operator given by $h(t,t') = W_{H}^\dagger(t) {G}_e(t,t^\prime) W_{H}(t^\prime) {G}_{g}^\da(t,t^\prime)$, with ${G}_{g}(t,t')\equiv \mathcal{T}[e^{-i\int_{t'}^t  g_{H}(s) ds}]$, ${G}_e(t,t')\equiv\mathcal{T}[e^{-i\int_{t'}^t  e_{H} (s) ds}]$, and $g_{H}(t)=\map H^{(a)}(t)\map$, $e_{H}(t)=\maq H^{(a)}(t)\maq$, and $W_{H}=\maq H^{(a)}(t)\map$. Note that the term $\int_0^t dt^\prime h(t,t')$ in Eq.~\eqref{TCLE} is responsible for leakage out of the ground state into excited states. Therefore, \emph{its vanishing constitutes a perturbative TCL condition for the transitionless evolution of the quantum system, and in particular represents a novel type of general adiabatic condition}. As we show in Appendix~\ref{sec:large-T}, the condition $\int_0^t dt^\prime h(t,t')=0$ is consistent with the usual adiabatic approximation in the sense that it is enforced in the limit as $T\to\infty$. However, as we shall illustrate, it is more general and can be applied to \emph{accelerate adiabaticity} in situations where the usual adiabatic condition does not apply, e.g., where the energy gap oscillates strongly, with a large average value.

In order to illustrate this approach in closed systems we revisit an example that was considered in Ref.~\cite{Jing:2014uq} without the TCL approach and only for the white noise model. We  consider a qubit prepared in the ground state $|E_0(0)\ra$ at $t=0$. By using Eq.~\eqref{TCLE}, it follows that the amplitude $c_0(t)$ obeys the TCL equation
\begin{equation}
\label{ME1}
\!\!\!\pa_t c_0(t)= \left[-\la E_0(t)|\dot{E}_0(t)\ra-\int_0^t dt^\prime \left. h(t,t') \right|_{11} \right] c_0(t)
\end{equation}
where $\left. h(t,t')\right|_{11}$ is the $(1,1)$ matrix element of $\map h(t,t') \map$, which is given by
\begin{eqnarray}
\left. h(t,t')\right|_{11} &=&\la E_0(t)|\dot{E}_1(t)\ra\la E_1(t^\prime)|\dot{E}_0(t^\prime)\ra  \\
&&\hspace{-1.8cm}\times \exp\left[\int_{t^\prime}^t dx \left(iE(x)+\la E_0|\dot{E}_0\ra - \la E_1|\dot{E}_1\ra \right)\right], \nonumber
\label{Pt}
\end{eqnarray}
with $E(x)\equiv E_0(x)-E_1(x)$.

Let us now show that fast, transitionless evolutions mimicking adiabaticity can be induced purely by manipulating $h(t,t')$. In this sense our approach provides an alternative to so-called shortcuts to adiabaticity (see, e.g., Ref.~\cite{Deffner:2014qv}). Toward this end, we consider a system subjected to a modification in its energy scale: $J_0\mapsto J=J_0+\Ga(t)$, which is the same as the control function $\Om(t)$ in the open quantum system case given in Sec.~\ref{example}. In order to show that our shortcut protocol is insensitive to the choice of control function, we consider three different choices for $\Gamma(t)$. In addition to the two (periodic rectangular pulse sequence and biased Poissonian continuous-time white noise) discussed in Sec.~\ref{example}, here we use another periodic pulse sequence but with a chaotic interaction intensity $\Psi_n=\Psi L_n$, where the dimensionless strength $L_n$ in different periods constitutes the logistic map $L_{n+1}=\mu(L_n-L_n^2)$, with $\mu=3.9$~\cite{Grebogi:1983ao}.

\begin{figure}[htbp]
\centering
\includegraphics[width=3.4in]{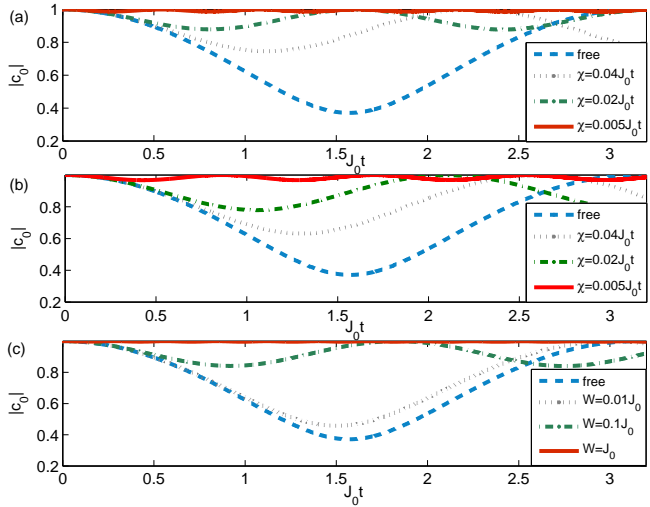}
\caption{ (Color online) Ground state amplitude $|c_0(t)|$ for a qubit in the presence of different control methods: (a) periodic rectangular pulse sequence; (b) periodic pulse sequence with chaotic strengths; (c) white noise. The blue curves in each panel represent the control-free evolution case. The parameters are chosen as $\Om=\om=5J_0$, so that the system is far from the adiabatic regime.}
\label{ca}
\end{figure}

First we consider a general time-dependent Hamiltonian for a qubit, which reads $H(t)=J\left(a\si^x+b\si^y+\frac{\om}{2}\si^z\right)$, where $J$ sets the energy scale, with $a$, $b$, and $\om$ describing magnetic fields in the $x$, $y$, and $z$ directions, respectively. Then the eigenstates of the Hamiltonian $H(t)$ can be expressed as $|E_0(t)\rangle=e^{-i\beta}\cos\gamma |\!\!\uparrow \rangle+ \sin\gamma \ket{\!\downarrow}$ and $|E_1(t)\ra=-e^{-i\beta}\sin\gamma\ket{\!\uparrow}+\cos\gamma\ket{\!\downarrow}$, where $\beta=\tan^{-1}(b/a)$ and $\gamma=\cos^{-1}\frac{k+\om}{\sqrt{2k^2+2k\om}}$ with $k\equiv\pm\sqrt{\om^2+4a^2+4b^2}$. We now consider a simple case given by $a=\cos(\Om t)$, $b=\sin(\Om t)$, with time-independent $\Om$ and $\om$. The transition operator for this model is
\begin{equation}
\left. h(t,t')\right|_{11}=\frac{\Om^2}{k^2}e^{i\Om(\sin^2\gamma)(t-t^\prime)}e^{i\int_{t^\prime}^t dx E(x)},
\end{equation}
where $E(x)=Jk=[J_0+\Gamma(x)]k$. Under free evolution (without modification of the original energy scale $J_0$), we obtain
\begin{equation}
\int_0^t dt^\prime \left. h(t,t')\right|_{11}=\frac{i\Om^2\left[1-e^{i(J_0k+\Om\sin^2\gamma)t}\right]}
{k^2(J_0k+\Om\sin^2\gamma)}.
\end{equation}
Thus, adiabaticity can be reached when $\Om$ approaches zero (indeed, here the conventional adiabatic condition is $\Om\ll\om$). Consider now a non-adiabatic regime where $\Om=\om=5J_0$. In Fig.~\ref{ca}, the blue curves depict the control-free evolution of $|c_0(t)|$, which oscillates from unity to $0.36$. In Fig.~\ref{ca}(a), we use pulse sequences with different periods to control $h(t,t')$, with the pulse strength $\Psi=0.01J_0$ and $\De/\chi=0.5$ fixed. By increasing the pulse repetition rate, $|c_0(t)|$ is made to approach unity at all times. In Fig.~\ref{ca}(b), the fixed pulse strength is replaced by chaotic pulses. Although the control effect is not as apparent as in Fig.~\ref{ca}(a), the same qualitative behavior is observed. In Fig.~\ref{ca}(c), we display the results of the biased Poissonian white noise case. The fluctuations of $|c_0(t)|$ are seen to be suppressed by increasing the noise strength $W$.

As a second example we now consider two coupled qubits, whose Hamiltonian is given by $H=J(d\si_1^{+}\si_2^-+d^*\si_1^-\si_2^{+}+B_1 \si_1^z+B_2 \si_2^z)$, where $d\equiv a-ib$ is a time-dependent parameter, $B_{1}=B+\om/4$ and $B_{2}=B-\om/4$, with $B$ playing the role of a noise parameter. If the system state is initialized as $\ket{\psi(0)}=\mu\ket{\!\!\uparrow\downarrow}+\nu\ket{\!\!\downarrow\uparrow}$ (with $|\mu|^2+|\nu|^2=1$), then $B$ acts on a time-dependent decoherence-free subspace~\cite{Zanardi:97c,Lidar:1998fk,Blume-Kohout:2008ec}, hence giving rise to no influence on the dynamics. Consequently, the effective Hamiltonian for this model can be written as $H_{\rm eff}=J[(d\si_1^+\si_2^-+h.c.)+\om(\si_1^z-\si_2^z)/4]$. The corresponding eigenstates of $H_{\rm eff}$ could also be expressed as those of a single qubit, via the mapping $|\!\!\uparrow\downarrow\ra\mapsto\ket{\!\uparrow}$ and $|\!\!\downarrow\uparrow\ra\mapsto\ket{\!\downarrow}$. Moreover, we also have in this case $\si_1^+\si_2^- \mapsto \si_{+}$ and $(\si_1^z-\si_2^z)/2 \mapsto \si_z$. We now let $a=t/T$, $b=0$, and $\om/2=1-t/T$. Then
\begin{equation}
\left. h(t,t')\right|_{11}=\frac{4}{T^2k^2(t)k^2(t^\prime)}e^{i\int_{t^\prime}^t dx [J_0+\Gamma(x)]k(x)},
\end{equation}
where $k(t)=2\sqrt{T^2-2tT+2t^2}/T$. Our goal is to realize the evolution from the eigenstate $|\!\!\uparrow\downarrow\ra$ of $H(0)=J_0(\si_1^z-\si_2^z)/2$ to the eigenstate $|\!\!\uparrow\downarrow\ra+|\!\!\downarrow\uparrow\ra$ of $H(T)=J_0(\si_1^+\si_2^-+h.c.)$.

\begin{figure}[htbp]
\centering
\includegraphics[width=3.4in]{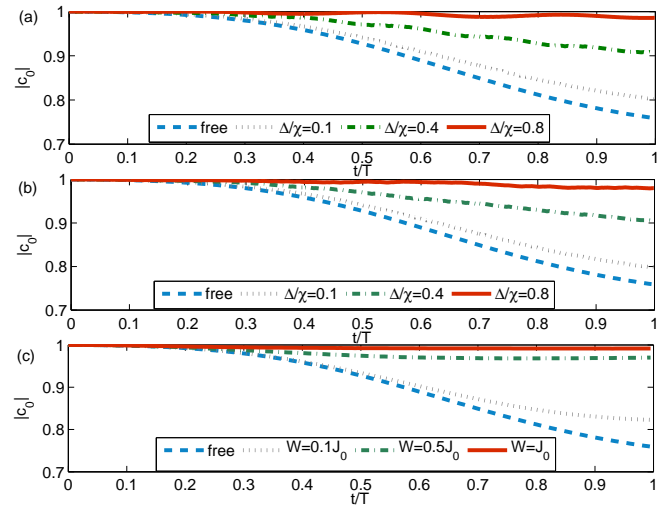}
\caption{ (Color online) Amplitude $|c_0(t)|$ for a coupled pair of qubits in the presence of different control methods: (a) periodic rectangular pulse sequence; (b) periodic pulse sequence with chaotic strengths; (c) white shot noise. The blue curves in each panel represent the control-free evolution of $|c_0(t)|$.}
\label{cb}
\end{figure}

In order to illustrate the non-adiabatic dynamics of the coupled system, we take $T=1/J_0$ and plot in Fig.~\ref{cb} the behavior of $|c_0(t)|$ as a function of the dimensionless time $t/T$. The blue curves depict the control-free case. Similar to Fig.~\ref{ca}, the other curves show the onset of the adiabaticity induced by ordered pulses [Fig.~\ref{cb}(a)], chaotic pulses [Fig.~\ref{cb}(b)] and white noise [Fig.~\ref{cb}(c)]. As can be seen, fast manipulation of $h(t,t')$ drives the system to the desired eigenstate of $H(T)$, i.e., $|\psi(T)\ra\approx|E_0(T)\ra$. Hence, it induces a transitionless evolution in a non-adiabatic regime.

\section{Discussion}
%%%%%%%%%%%%%%%%
%\noindent\textbf{Discussion}
%%%%%%%%%%%%%%%%

We have employed the Feshbach P-Q partitioning and the Nakajima-Zwanzig projection techniques to derive a general one-component projected time-convolutionless equation for open quantum systems described by time-dependent master equations: Eq.~\eqref{eq:TCL-result}. By choosing the projection to the ground state manifold, we were able to identify new conditions for ground-state dynamics in open quantum systems. These conditions provide an alternative design for shortcuts to adiabaticity beyond the transitionless-tracking algorithm~\cite{Rice1,Rice2,Berry:2009kx,
Torrontegui:13,Campo:2012il,Campo:2013ix,Deffner:2014qv,Papoular}, that does not require the addition of (typically highly non-local) counter-diabatic driving to mimic adiabatic behavior. Specifically, we have demonstrated the onset of transitionless evolutions induced by fast control through, e.g., periodic rectangular pulses, white noise, and chaotic signals. The insensitivity of the success of the ground state tracking protocol to the details of the control sequence shows that the protocol is highly robust, in particular to noise. Moreover, the evolution of the system under our control protocols not only connects the two eigenstates of the initial and final Hamiltonians, but also adheres to an adiabatic path. Therefore there is no need for a precise design of counter-diabatic driving or time-dependent confining potentials. Similarly to dynamical decoupling protocols for protecting adiabatic evolution from the effects of coupling to the bath, our methodology also involves fast controls, but does not require encoding. These results imply the existence of versatile schemes for shortcuts to adiabaticity in both closed or open quantum systems, with direct impact on the physical implementation of quantum control protocols, in particular in open system adiabatic quantum computation and quantum annealing. We expect that  eigenstate tracking techniques for open system will play a key role in this context, as well as in  quantum thermodynamics~\cite{Hoffmann:2011ts,Deng:2013xu,Campo:2014rf,Stefanatos:2014bl} and dissipative quantum state engineering~\cite{diehl2008quantum,verstraete2009quantum,Huelga:2012zf}.

%\vspace{0.7cm}

\section*{Acknowledgements}

J.J, D.W.L., and L.A.W. acknowledge support from the NSFC No. 11575071, the Basque Government (grant IT472-10), and the Spanish MICINN (No. FIS2012-36673-C03- 03). M.S.S. acknowledges support from CNPq / Brazil (No. 304237/2012-4) and the Brazilian National Institute for Science and Technology of Quantum Information (INCT-IQ). D.A.L.'s work was supported under under ARO MURI Grant No. W911NF-11-1-0268 and ARO grant number W911NF-12-1-0523.

%%%%%%%%%%%%%%%%
%\noindent\textbf{Methods}\\
\appendix
\label{sec:methods}
%%%%%%%%%%%%%%%%

\section{Formal solution to Eq.~\eqref{Oqqphi}}
\label{sec:check-sol}
We verify that Eq.~\eqref{Oqphi} is the solution to Eq.~\eqref{Oqqphi} as claimed in the main text. Indeed, upon differentiation of Eq.~\eqref{Oqphi}, use of $\pa_t \mathcal{G}(t,t^\prime) = \maq {\cal L}_I(t) \mathcal{G}(t,t^\prime)$ and $\mc{G}(t,t) = \openone$ we obtain:
\begin{align}
\pa_t\maq\kett{\chi(t)} &= \maq {\cal L}_I(t) \left[ \mc{G}(t,0) \maq \kett{\chi(0)} + \mc{P}\kett{\chi(t)}  \right.\notag \\
&\left. \,\,\,\,\,\,+  \int_0^t \mc{G}(t,t') \mc{Q} \mc{L}_I(t') \mc{P}\kett{\chi(t')} dt' \right] \notag \\
& = \maq {\cal L}_I(t) \mc{P}\kett{\chi(t)} + \maq {\cal L}_I(t) \maq \kett{\chi(t)} ,\notag
\end{align}
as required by Eq.~\eqref{Oqqphi}, where in the second line we use Eq.~\eqref{Oqphi} again.

%%%%%%%%%%%%%%%%%%%%%%%%%%%
%\noindent\textbf{Adiabatic perturbative expansion in open systems.}
\section{Adiabatic perturbative expansion in open quantum systems}
\label{sec:ad-pert}
%%%%%%%%%%%%%%%%%%%%%%%%%%%
Here we present the adiabatic perturbation theory steps described in the main text. Our analysis will also give the conditions for the invertibility of $\left[ 1 - \Sigma(t) \right]$, which is needed for the TCL dynamical equation, Eq.~\eqref{eq:TCL-result}. As in the main text, we define the normalized time $s=t/T$, where $s\in[0,1]$. Changing variables in ${\cal H}_I(t)$ [Eq.~\eqref{eq:I-pic}] yields ${\cal H}_I(t) \mapsto {\cal H}_I(s,T) $, with
\begin{equation}
 {\cal H}_I(s,T) = {\cal U}_0^\da (s,T) \left[ W_{\cal H}(s,T) + W_{\cal H}^\da(s,T) \right] {\cal U}_0 (s,T) .
\end{equation}
The matrix elements of the operator ${\cal H}_I(s,T)$ can be written as $\left[ {\cal H}_I(s,T) \right]_{mn} = \left[ {\cal H}^W_I(s,T) \right]_{mn} + \left[ {\cal H}^W_I(s,T) \right]_{mn}^*$, with $\left[ {\cal H}^W_I(s,T) \right]_{mn} = \sum_p  [{\cal U}_0^{\da}]_{mp} [{W}_{\cal H}]_{p0} [{\cal U}_0]_{0n}$. From Eq.~\eqref{H-ad-frame}, we then obtain
\begin{equation}
\left[ {\cal H}^W_I(s,T) \right]_{mn} = -\frac{i}{T} \sum_p {\cal{O}}_{mnp}(s,T) e^{-i T \int_0^s dx \, \Omega_{0p}(x)},
\label{eq:37}
\end{equation}
where ${\cal{O}}_{mnp}(s,T) = [{\cal U}_0^{\da}]_{mp}  \langle \langle \Phi_p | \frac{\partial}{\partial s} | \Phi_0 \rangle \rangle [ {\cal U}_0]_{0n}$. We are now ready to apply perturbation theory in terms of $1/T$, which will appear whenever $\left[ {\cal H}^W_I(s,T) \right]_{mn}$ is integrated. First note that, by Eq.~\eqref{eq:37},  $i \int_0^t dt^\prime \left[ {\cal H}^W_I(t^\prime)\right]_{mn} = i T \int_0^s ds^\prime \left[ {\cal H}^W_I(s^\prime,T)\right]_{mn} = \sum_p I_{mnp} (s,T)$, with
\begin{equation}
I_{mnp}(s,T) =  \int_0^s ds^\prime {\cal{O}}_{mnp}(s^\prime,T) e^{ -i T \int_0^{s^\prime} dx \, \Omega_{0p}(x)} .
\label{Imnp}
\end{equation}
So far $T$ cancels out. However, integration by parts will yield $1/T$ due to the presence of $T$ in the dynamical phase term. Indeed, letting $u={\cal{O}}_{mnp}(s^\prime,T)/\Omega_{0p}(s^\prime)$ and $dv = \Omega_{0p}(s^\prime)\exp[ -i T \int_0^{s^\prime} dx \, \Omega_{0p}(x) ]ds'$, we obtain after integrating by parts with $\int u dv = uv - \int v du$:
\begin{eqnarray}
\label{ibp}
I_{mnp}(s,T)&= \frac{i}{T} \left[\frac{{\cal{O}}_{mnp}(s^\prime,T)}{\Omega_{0p}(s^\prime)}e^{-i T \int_0^{s^\prime}\Omega_{0p}(x) dx}\right]\bigg|_{s^\prime=0}^{s^\prime=s}  \notag  \\
\label{eq:39b}
&-\frac{i}{T} \int_0^s {\cal{O}}^{(1)}_{mnp}(s^\prime,T) e^{-i T\int_0^{s^\prime}\Omega_{0p}(x) dx} ,
\end{eqnarray}
where ${\cal{O}}^{(1)}_{mnp}(s^\prime,T) \equiv {\frac{\partial}{\partial{s'}} \frac{{\cal{O}}_{mnp}(s^\prime,T)}{\Omega_{0p}(s^\prime)}}$. Note that the boundary term~\eqref{eq:39b} is of order $1/T$. Comparing Eq.~\eqref{Imnp} and Eq.~\eqref{eq:39b}, it is clear that a second integration by parts, accomplished by letting $u ={\cal{O}}^{(1)}_{mnp}(s^\prime,T)$ and $dv$ as above, will produce a new boundary term of order $O(T^{-2})$ from the $v$ term. Repeating the integration by parts, we can write a perturbation expansion of the form $I_{mnp}(s,T) =\sum_{k=1}^\infty T^{-k} I^{(k)}_{mnp}(s)$, where $I^{(k)}_{mnp}(s)$ is the $k^{\rm th}$ boundary term resulting from the $k^{\rm th}$ integration by parts. In summary, we can perform the identification
\begin{eqnarray}
i T \int_0^s ds^\prime \left[ {\cal H}^W_I(s^\prime,T)\right]_{mn} &=&    \\
&&\hspace{-3.5cm} \frac{i}{T} \int_0^s ds^\prime \left[\overline{{\cal H}}^{W}_I(s^\prime ,T)\right]_{mn} + O(T^{-2}) , \nonumber
\end{eqnarray}
with
\begin{eqnarray}
\left[\overline{{\cal H}}^{\, W}_I(s^\prime,T)\right]_{mn} \hspace{-0.1cm}&=& \hspace{-0.1cm}  \\
&&\hspace{-2.5cm}\frac{d}{ds^\prime}
\sum_{p}  \left[\frac{{\cal{O}}_{mnp}(s^\prime,T)}{\Omega_{0p}(s^\prime)}
e^{-iT\int_0^{s^\prime}\Omega_{0p}(x) dx}\right]\bigg|_{s^\prime=0}^{s^\prime=s} \nonumber
\end{eqnarray}
Similarly to ${\cal H}_I(s,T)$, the integration by parts of the decoherence contribution ${\cal D}_I(s,T)$ also exhibits a leading order $T^{-1}$. This is again due to the presence of the dynamical phase $e^{ -i [\Theta_l(s)-\Theta_k(s)]}$ [see Eq.~\eqref{Lind-af2}].

With these results at hand, we now address the invertibility of $\left[ 1 - \Sigma(t) \right]$. Concerning $\Sigma(t)$, we obtain from Eq.~\eqref{eq:Sigma} that its $1/T$-expansion after the change of variables $\Sigma(t) \mapsto \Sigma(s,T)$ yields
\begin{eqnarray}
\Sigma(s,T) \hspace{-0.15cm}&=&\hspace{-0.1cm} T \int_0^s \mc{G}(s,s',T)\mc{Q}\mc{L}_I(s',T)\mc{P}\mc{V}_I^\dagger(s,s',T) ds' \notag\\
\hspace{-0.15cm}&=& \hspace{-0.1cm}T \int_0^s \hspace{-0.1cm}[\openone + O(\mc{Q} \mc{L}_I)] \mc{Q}\mc{L}_I(s',T)\mc{P}[\openone + O(\mc{L}_I)] ds'  \notag \\
\hspace{-0.15cm}&=&\hspace{-0.1cm} T \int_0^s \hspace{-0.1cm} \mc{Q}\mc{L}_I(s',T)\mc{P} ds' + O(\mc{L}_I^2) \notag \\
\hspace{-0.15cm}&=&\hspace{-0.1cm}T \int_0^s  \hspace{-0.1cm}\mc{Q}[-i \mc{H}_I(s',T) \hspace{-0.1cm}+ \hspace{-0.1cm}\mc{D}_I(s',T)]\mc{P} ds' + O(\mc{L}_I^2). \nonumber \\
\end{eqnarray}
Since $\int_0^s ds' \mc{H}_I(s',T) = O(1/T^2)$, as we have already shown,  this contribution vanishes in the large $T$ limit, which ensures the invertibility of $[\openone - \Sigma(s,T)]$ for closed systems. The contribution of the decoherence superoperator needs not vanish, since $\mc{D}_I(s',T) = O(1/T)$, as we have also already shown. Therefore the leading order contribution to $\Sigma(s,T)$ is a constant ($T$-independent) term. However, we may assume that the contribution of the decoherence superoperator $\mc{D}_I$ is perturbative due to the weak system-bath coupling, as we did after Eq.~\eqref{TCLs}. It is then consistent to assume that $T \int_0^s  \mc{Q}\mc{D}_I(s',T)\mc{P} ds'$ is a perturbation of the identity, which ensures the invertibility of $[\openone - \Sigma(s,T)]$ for open systems in the weak-coupling limit.

%%%%%%%%%%%%%%%%%%%%%%%%%%%%%%%%%%%%%%%%%%%%%%%%%%
\section{Open-system example: derivation of Eqs.~\eqref{eq:hf} and~\eqref{Fiall}}
\label{sec:check-eqhf}
%%%%%%%%%%%%%%%%%%%%%%%%%%%%%%%%%%%%%%%%%%%%%%%%%%
Let us consider here the two-level open system described in Subsection~\ref{example}. We prepare the system in the excited state $|E_{-}(0)\rangle$ and are interested in the fidelity ${\cal F}\equiv\sqrt{\la E_-(T)|\rho(T)|E_-(T)\ra}$. For the computation of the exact fidelity, we further change variables in the master equation~\eqref{exactME} through $\rho \rightarrow \rho_I (t) = U(t)  \rho(t)  U^\dagger(t)$, with $U(t)  = \exp \left[ i \int_0^t dt' \left(J(t') + {\mathcal S}(t')\right) \sigma_z(t')\right]$. Then, we obtain
\begin{equation}
\pa_t\rho_I(t)\hspace{-0.1cm}=\hspace{-0.1cm}\ka (t) \left[\sigma_-(t)\rho_I(t)\si_+(t)-\frac{1}{2}\left\{\sigma_+(t)\sigma_-(t),\rho_I(t) \right\}\right].
\label{exactMEI}
\end{equation}
The matrix element of $\rho_I(t)$ associated with the fidelity is $\rho_I^{-}(t) \equiv \langle E_{-}(t)|\rho_I | E_-(t)\rangle$. In particular, notice that ${\cal F} = \sqrt{\la E_-(T)|\rho(T)|E_-(T)\ra} = \sqrt{\rho_I^{-}(t)}$, since $U^\dagger(T)|E_-(T)\rangle = \exp \left[ -i \int_0^T dt' \left(J(t') + {\mathcal S}(t')\right) \right] |E_-(T)\rangle$. To obtain $\rho_I^{-}(t)$, we use Eq.~(\ref{exactMEI}), which yields
\begin{equation}
\pa_t \rho_{I}^{-}(t) = - \ka(t) \rho_{I}^{-}(t) .
\end{equation}
Hence, as provided by Eq.~(\ref{eq:hf}), the exact fidelity reads
\begin{equation*}
{\cal F}(T)=\exp\left[-\frac{1}{2}\int_0^Tdt\ \ka(t)\right].
\end{equation*}

In order to obtain the perturbative TCL fidelity ${\cal F}\equiv\sqrt{\la E_-(T)|\rho(T)|E_-(T)\ra}=\sqrt{|r_0(T)|}$, with $r_0(T)$ denoting here the excited state population and being provided by Eq.~\eqref{de-open}, we have to determine the Hamiltonian ${\cal H}^{(a)}(t)$ and the decoherence ${\cal D}^{(a)}(t)$ superoperators in the adiabatic frame. Then, by computing ${\cal H}^{(a)}(t)$ and ${\cal D}^{(a)}(t)$ according to Eqs.~\eqref{H-ad-frame} and \eqref{Lind-af2}, respectively, we obtain
\begin{equation}\label{appH}
{\cal H}^{(a)}=\frac{\pi}{4T}\left(\begin{array}{cccc} 0 & -ie^{i\bar{J}(t)} & -ie^{-i\bar{J}(t)} & 0 \\  ie^{-i\bar{J}(t)} & 0 & 0 & -ie^{-i\bar{J}(t)} \\ ie^{i\bar{J}(t)} & 0 & 0 & -ie^{i\bar{J}(t)}\\ 0 & ie^{i\bar{J}(t)} & ie^{-i\bar{J}(t)}& 0 \end{array}\right),
\end{equation}
where $\bar{J}(t)\equiv2\int_0^tdt'J(t')$, and
\begin{equation}\label{appD}
{\cal D}^{(a)}=\left(\begin{array}{cccc} -\ka(t) & 0 & 0 & 0 \\  0 & 2i{\cal S}(t)-\frac{\ka(t)}{2} & 0  &  0 \\  0 &  0 & -2i{\cal S}(t)-\frac{\ka(t)}{2} &  0 \\ \ka  & 0  & 0  & 0 \end{array}\right).
\end{equation}
According to Eqs.~\eqref{geWDa} and \eqref{geWDb}, we can determine the superoperators $g_{\cal H}$, $e_{\cal H}$, and $W_{\cal H}$ through the decomposition of Eq.~\eqref{appH}. In particular, we obtain $g_{\cal H}= \mathbb{0}_{N^2}$. Then, from Eq.~\eqref{GgGe}, we get ${\cal G}_g(t,t')=\mathbb{1}_{N^2}$. Similarly, we can also find out $g_{\cal D}$, $e_{\cal D}$, $W_{\cal D}$, and $V_{\cal D}$ through the decomposition of Eq.~\eqref{appD}. In particular, we obtain $g_{\cal D}=-\ka \map$ and $V_{\cal D}=\mathbb{0}_{N^2}$. Therefore, Eqs.~\eqref{gopen} and \eqref{fopen} can be explicitly written as
\begin{eqnarray}
\left.h(t, t')\right|_{11}\hspace{-0.1cm}&=&\hspace{-0.1cm}\left(\frac{\pi}{4T}\right)^2
[0,-ie^{2i\ti{J}t}, -ie^{-2i\ti{J}t}, 0] \notag \\
&& \hspace{1cm}{\cal G}_e(t,t^\prime) [0,ie^{-2i\ti{J}t'}, ie^{2i\ti{J}t'}, 0]^T \notag \\
&\approx& \frac{\pi^2}{8T^2}\cos[2\ti{J}(t-t')]
\label{app:gf-opena}
\end{eqnarray}
and
\begin{eqnarray}
\left.f(t, t')\right|_{11}\hspace{-0.1cm} &=&\hspace{-0.1cm} i \frac{\pi}{4T}[0,-ie^{2i\ti{J}t}, -ie^{-2i\ti{J}t}, 0] \notag \\
&& \hspace{0.7cm}{\cal G}_e(t,t^\prime) [0,0, 0, \ka(t')]^T \notag \\
&\approx&-\frac{\pi^2\ka(t')}{16T^2\ti{J}}\sin[2\ti{J}(t-t')] .
\label{app:gf-openb}
\end{eqnarray}
To obtain Eqs.~\eqref{app:gf-opena} and~\eqref{app:gf-openb}, we approximated the average gap $\ti{J}$ by a constant and considered the expansion
\begin{equation}
{\cal G}_e(t,t')={\cal T}\left[e^{-i\int_{t'}^te_{\cal H}(x)dx}\right] \approx{\cal I}-i\int_{t'}^t \hspace{-0.15cm}e_{\cal H}(x)dx,
\end{equation}
where
\begin{equation}
-i\int_{t'}^t \hspace{-0.15cm} e_{\cal H}(x)dx =\frac{\pi}{8T\ti{J}}
\left(\begin{array}{cccc} 0 & 0 & 0 & 0  \\ 0& 0 & 0 & -A-iB\\ 0& 0 & 0 & -A+iB\\ 0&  A-iB & A+iB& 0\end{array}\right),
\end{equation}
with $A\equiv\sin2\ti{J}t-\sin2\ti{J}t'$ and $B\equiv\cos2\ti{J}t-\cos2\ti{J}t'$. Note that the adiabatic expansion of the propagator ${\cal G}_e(t,t')$ provides the leading order $(1/T^2)$ for $f(t,t')$ in Eq.~\eqref{app:gf-openb}. Moreover, observe that, by treating $\ka(t)$ as a perturbative parameter (weak-coupling regime), we have $||f(t,t')||\ll||h(t,t')||$. We can then disregard the contribution for the fidelity from $f(t,t')$ with respect to $h(t,t')$. Hence, by using Eq.~\eqref{app:gf-opena}, we obtain the fidelity through the perturbative TCL master equation~\eqref{de-open}, which reads
\begin{eqnarray*}
&& {\cal F}_{\rm TCL}(T)\\ &\approx& \exp\left[-\frac{1}{2}\int_0^T
dt \ka(t)-\frac{1}{2}\int_0^T dt\int_0^t dt' \left. h(t,t') \right|_{11}\right], \\ &=&\exp\left[-\frac{1}{2}\int_0^T dt \ka(t)+\left(\frac{\pi}{8}\right)^2\frac{\left(\cos2\ti{J}T-1\right)}{(\ti{J}T)^2}\right].
\end{eqnarray*}

%%%%%%%%%%%%%%%%%%%%%%%%%%%%%%%%%%%
%\noindent\textbf{Adiabaticity in closed systems: the large $T$ limit.}
\section{Adiabaticity in closed systems: the large $T$ limit}
\label{sec:large-T}
%%%%%%%%%%%%%%%%%%%%%%%%%%%%%%%%%%%
In order to show that the transitionless condition $\int_0^t dt^\prime h(t,t')=0$ includes, as a particular case, the ordinary adiabatic condition for a large total evolution time $T$, we reintroduce the normalized time $s$. Then, by rewriting Eq.~\eqref{TCLE} in terms of $s$, we obtain
\begin{equation}
\pa_s c_0(s)=\left[- \la E_0(s) | \pa_s |E_0(s) \ra - T^2 \int_0^s \hspace{-0.15cm} ds^\prime g(s,s^\prime)\right]c_0(s),
\end{equation}
where
\begin{eqnarray}
g(s,s^\prime) &=& \frac{1}{T^2} \sum_{p,q=1}^{N-1} [W_H^\dagger(s)]_{1p} \, [G_e(s,s^\prime)]_{pq} \, \nonumber \\
&&\hspace{1.3cm}[W_H(s^\prime)]_{q1} \, [G_g^\da(s,s^\prime)]_{11}.
\end{eqnarray}

However, note that the integration over $g(s,s^\prime)$ will only affect the terms $[G_e(s,s^\prime)]_{pq} \, [W_H(s^\prime)]_{q1} \, [G_g^\da(s,s^\prime)]_{11}$. Then, the integral $[I(s)]_{pq}$ for an individual term $(p,q)$ in the sum can be written as
\begin{eqnarray}
[I(s)]_{pq} &\equiv& \int_0^s ds^\prime  [\mathcal{T} e^{iZ(s,s^\prime)}]_{pq} \nonumber \\
&&\hspace{-0.6cm}\left[ -i \la E_q(s^\prime) | \pa_{s^\prime} |E_0(s^\prime) \ra e^{i T \int_0^{s^\prime} dx  \omega_{q0}(x) } \right],
\end{eqnarray}
where $\omega_{q0}(x) \equiv E_q(x) - E_0(x)$ and  $[\mathcal{T} e^{iZ(s,s^\prime)}]_{pq} \equiv [G_e(s,s^\prime)]_{pq}\, [G_g^\da(s,s^\prime)]_{11}$. Then, integrating by parts, we obtain
\begin{eqnarray*}
[I(s)]_{pq}&=&-e^{iT\int_0^{s}\omega_{q0}(x) dx}  \frac{\la E_q(s) | \pa_s |E_0(s) \ra}{T\,\omega_{q0}(s)}\\ &\times& \left[ \mathcal{T} e^{iZ(s,s')}\right]_{pq} \bigg|_0^s+\int_0^s ds' e^{i T \int_0^{s'}\omega_{q0}(x) dx} \\ &\times&  \frac{\pa}{\pa s^\prime} \left[ [\mathcal{T} e^{iZ(s,s^\prime)}]_{pq}  \frac{\la E_q(s^\prime) | \pa_{s^\prime} |E_0(s^\prime) \ra}{T\,\omega_{q0}(s^\prime)}\right] .
\end{eqnarray*}

Now, by using the Riemann-Lebesgue lemma~\cite{Ryzhik,Churchill:Book}, we can obtain a vanishing integral $[I(s)]_{pq}$ by imposing that
\begin{eqnarray}
T &\gg & \max_{s,q} \left| \frac{\la E_q(s) | \pa_s |E_0(s) \ra}{\omega_{q0}} \right| \nonumber \\
&=&  \max_{s,q} \left| \frac{\la E_q(s) | [\pa_s H(s)] |E_0(s) \ra}{\omega_{q0}^2} \right| \,\,\,\,(q \ne 0) . \label{ord-adiab}
\end{eqnarray}
Note that Eq.~\eqref{ord-adiab} is exactly the ordinary adiabatic condition~\cite{Messiah:book}. It has been obtained here from the transitionless condition $\int_0^t dt^\prime h(t,t')=0$ by requiring a total evolution time $T$ that is large in comparison with the inverse of the minimum instantaneous energy gap. However, note that the transitionless condition is more general than Eq.~\eqref{ord-adiab}. In particular, as we have shown, it may achieve an acceleration of the adiabatic path by inducing a fast oscillating gap.

%\bibliographystyle{apsrev4-1}
%\bibliography{refs}

\end{document}